\documentclass[journal=jacsat,manuscript=article]{achemso}
\usepackage{amsmath}
\usepackage{amssymb}   % need both to use \gtrsim command (and symilar mathematical symbols)
\usepackage{hyperref}  % for clickable references
\usepackage{mhchem}    % for chemical formulas
\usepackage{xcolor}    % to use different colors (\textcolor{color}{text} command)
\SectionNumbersOn

\newcommand{\beginsupplement}{%
	\setcounter{table}{0}
	\renewcommand{\thetable}{S\arabic{table}}%
	\setcounter{figure}{0}
	\renewcommand{\thefigure}{S\arabic{figure}}%
	\renewcommand{\thesection}{S\arabic{section}}%
}

\author{Ivan Villani}
\email{ivan.villani@sns.it}
\affiliation{NEST, Istituto Nanoscienze-CNR and Scuola Normale Superiore, Piazza San Silvestro 12, 56127 Pisa, Italy}

\author{Matteo Carrega}
\affiliation{CNR-SPIN, Via Dodecaneso 33, 16146 Genova, Italy}

\author{Alessandro Crippa}
\affiliation{NEST, Istituto Nanoscienze-CNR and Scuola Normale Superiore, Piazza San Silvestro 12, 56127 Pisa, Italy}

\author{Elia Strambini}
\affiliation{NEST, Istituto Nanoscienze-CNR and Scuola Normale Superiore, Piazza San Silvestro 12, 56127 Pisa, Italy}

\author{Francesco Giazotto}
\affiliation{NEST, Istituto Nanoscienze-CNR and Scuola Normale Superiore, Piazza San Silvestro 12, 56127 Pisa, Italy}

\author{Vaidotas Mišeikis}
\affiliation{Center for Nanotechnology Innovation, Laboratorio NEST, Istituto Italiano di Tecnologia, Piazza San Silvestro 12, 56127 Pisa, Italy}

\author{Camilla Coletti}
\affiliation{Center for Nanotechnology Innovation, Laboratorio NEST, Istituto Italiano di Tecnologia, Piazza San Silvestro 12, 56127 Pisa, Italy}

\author{Fabio Beltram}
\affiliation{NEST, Istituto Nanoscienze-CNR and Scuola Normale Superiore, Piazza San Silvestro 12, 56127 Pisa, Italy}

\author{Kenji Watanabe}
\affiliation{Research Center for Electronic and Optical Materials, National Institute for Materials Science, 1-1 Namiki, Tsukuba 305-0044, Japan}

\author{Takashi Taniguchi}
\affiliation{Research Center for Materials Nanoarchitectonics, National Institute for Materials Science, 1-1 Namiki, Tsukuba 305-0044, Japan}

\author{Stefan Heun}
\email{stefan.heun@nano.cnr.it}
\affiliation{NEST, Istituto Nanoscienze-CNR and Scuola Normale Superiore, Piazza San Silvestro 12, 56127 Pisa, Italy}

\author{Sergio Pezzini}
\email{sergio.pezzini@nano.cnr.it}
\affiliation{NEST, Istituto Nanoscienze-CNR and Scuola Normale Superiore, Piazza San Silvestro 12, 56127 Pisa, Italy}

\title{Quasi-$\Phi_0$-periodic supercurrent at quantum Hall transitions}

\begin{document}
	
\begin{abstract}

	The combination of superconductivity and quantum Hall (QH) effect is regarded as a key milestone in advancing topological quantum computation in solid-state systems. 
	Recent quantum interference studies suggest that QH edge states can effectively mediate a supercurrent across high-quality graphene weak links. In this work we report the observation of a supercurrent associated with transitions between adjacent QH plateaus, where transport paths develop within the compressible two-dimensional bulk. We employ a back-gated graphene Josephson junction, comprising high-mobility CVD-grown graphene encapsulated in hexagonal Boron Nitride (hBN) and contacted by Nb leads. Superconducting pockets are detected persisting beyond the QH onset, up to $2.4$~T, hence approaching the upper critical field of the Nb contacts. We observe an approximate $\Phi_0=h/2e$ periodicity of the QH-supercurrent as a function of the magnetic field, indicating superconducting interference in a proximitized percolative phase. These results provide a promising experimental platform to investigate the transport regime of percolative supercurrents, leveraging the flexibility of van der Waals devices.

\end{abstract}


Quantum computing hardware resilient to environmental disturbances can be engineered based on specific topological phases of matter, which allow non-local information storage and manipulation via quasiparticle exchange in real space \cite{Lindner2012,Stern2013}. This paradigm can be realized in systems where QH states are interfaced with s-wave superconductors \cite{Giazotto2005}, supporting non-Abelian anyons such as Majorana zero modes \cite{SanJose2015} and parafermions \cite{Clarke2013,Alicea2016}. These platforms are regarded as fundamental building blocks for topological quantum computation, paving the way toward a universal set of quantum gates \cite{Sergey2005, Nayak2008}. Consequently, significant efforts have been dedicated to overcome the intrinsic challenges of proximitizing QH states, both in III-V semiconductors \cite{Wan2015, Zhi2019, Guiducci2019, Hatefipour2022, Hatefipour2024, Akiho2024} and graphene \cite{Amet2016,Lee2017, Zhao2020, Wang2021, Gul2022, Vignaud2023, Barrier2024, Zhao2024, jang2025arxiv}.

hBN-encapsulated graphene devices are considered an ideal experimental platform, owing to their ability to support highly-transparent one-dimensional contacts \cite{Wang2013, Calado_2015} which facilitate Andreev reflection both in the integer \cite{Lee2017} and fractional \cite{Gul2022} QH regime. Additionally, these devices enable coherent edge-state propagation over micrometers \cite{Ronen2021, Deprez2021}, further enhancing their suitability for quantum transport experiments.
The first observation of a supercurrent (SC) in the QH regime by Amet \textit{et al.} \cite{Amet2016} was made possible by Josephson junctions with this configuration. The SC was attributed to the coupling between chiral Andreev Edge States (CAES), hybrid electron-hole modes propagating along the graphene-superconductor interface \cite{Zhao2020}, and QH edge states, forming a closed loop. The resulting chiral supercurrent, subject to Aharonov-Bohm interference, is expected to display a $2\Phi_0=h/e$ periodicity as a function of the applied magnetic field \cite{Alavirad2018,Liu2017}, implying that it oscillates at half the frequency of a conventional non-chiral Josephson current. However, an unexpected $\Phi_0=h/2e$ periodicity was reported in Ref.~\citenum{Amet2016} for a SC in the QH regime. 
Subsequent studies \cite{Draelos2018,Seredinski2019} suggested that counterpropagating channels capable of independently carrying a SC emerge due to charge accumulation at etched graphene edges \cite{Aharon_Steinberg_2021}, effectively forming a superconducting quantum interference device (SQUID).
Recent calculations show that the $\Phi_0$-periodicity could also arise from finite coupling between the CAES wavefunctions in the short junction limit \cite{Sun2024}. By miniaturizing device edges (width $W$ and length $L$ $<330$ nm) to enhance phase coherence \cite{Kurilovich2023}, Vignaud \textit{et al.} were able to observe a $2\Phi_0$-periodic chiral QH-SC at filling factor $\nu=2$ \cite{Vignaud2023}. More recently, Barrier \textit{et al.} \cite{Barrier2024} reported a QH-SC carried by one-dimensional domain walls within the bulk of minimally twisted bilayer graphene (hence not relying on physical edges) persisting up to magnetic fields very close to the critical field of the superconducting contacts. Altogether, these results highlight the crucial influence of sample boundaries, including both  graphene-superconductor and graphene-vacuum interfaces, in QH-SC coupling. However, charge transport in the QH regime can also involve bulk extended states at transitions between different QH plateaus \cite{WeiTsui1988}. In such cases, two-dimensional systems undergo a localization-delocalization transition \cite{Pruisken1988}, which was modeled (semi-classically) in terms of percolation \cite{Trugman1983} through a network of QH droplets generated by long-range potential fluctuations \cite{Shapiro1986}.

In this work we present experimental evidence of QH-SC in Nb-contacted encapsulated graphene, persisting up to ${B=2.4}$~T, close to the Nb upper critical field (${B_c=3.2}$~T). 

The SC is observed at QH plateau-plateau transitions and exhibits an approximate $\Phi_0$ periodicity as a function of magnetic field. Our findings suggest a mechanism analogous to low-field Fraunhofer pattern in planar Josephson junctions, enabled by the formation of percolative bulk channels. The interference involves different areas throughout the device, governed by the interplay of applied electromagnetic fields and doping near the graphene-superconductor interface.

\section*{Results and discussion}

\subsection*{Device properties at zero and low magnetic field}

\begin{figure}[H]
	\centering
	\includegraphics[width=\linewidth]{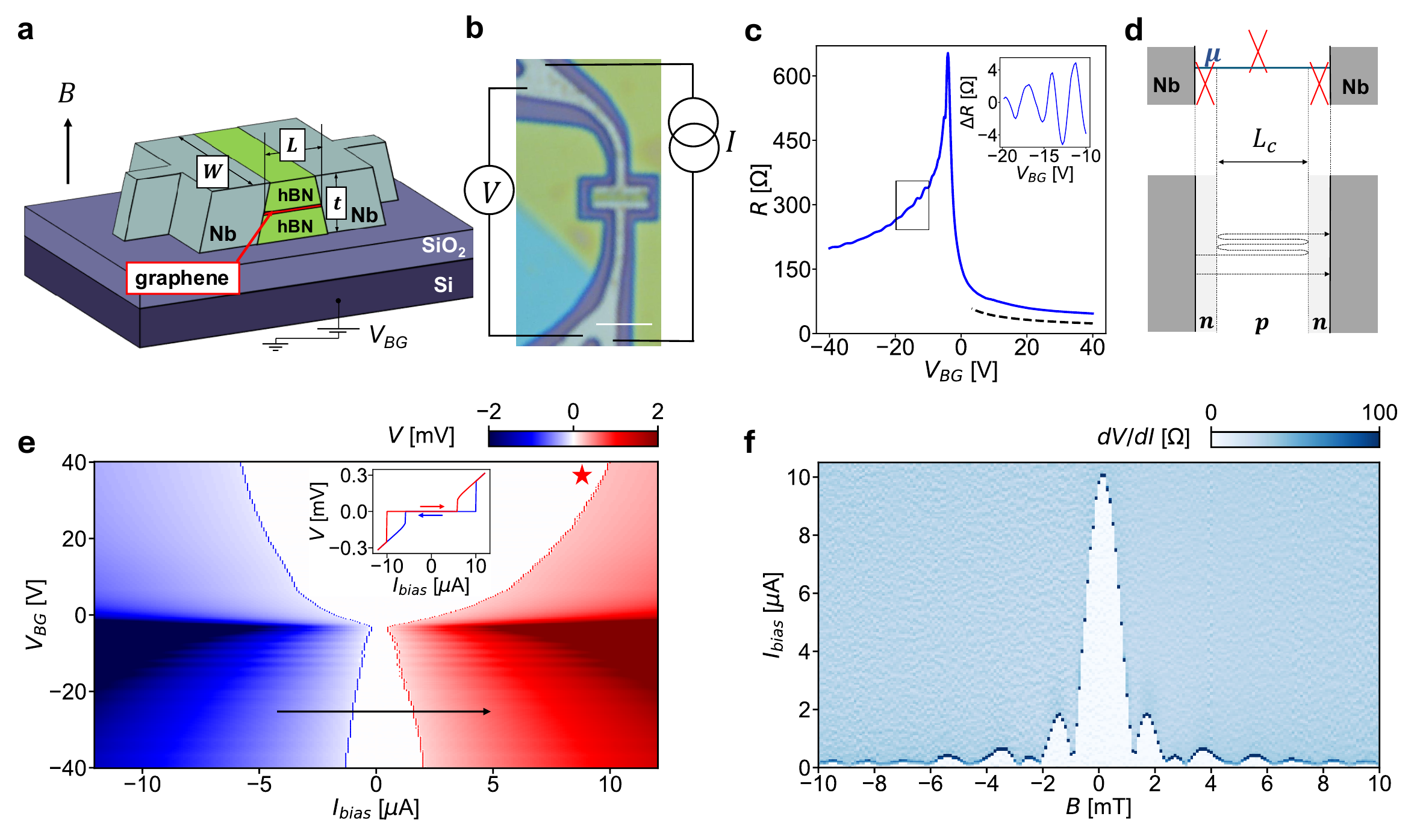}
	\caption{\label{fig1} ($\mathbf{a}$) 3D schematics of the device layout and gating configuration. When applied, the magnetic field $B$ is orthogonal to the substrate plane. Device dimensions: $L=400$~nm, $W=3$~$\mu$m, $t=60$~nm. (schematics is not to scale). ($\mathbf{b}$) Optical microscopy image of the junction. Scalebar: $3\;\mu$m. The four-probe measurement layout is indicated: a current $I$ (combining DC and AC components as specified in the main text) is applied to the junction, and the voltage drop $V$ is measured. ($\mathbf{c}$) Representative backgate sweep: sample resistance $R=V/I$ as a function of the backgate voltage $V_{BG}$. $T = 4.2$~K, $B=0$~T. The normal state resistance is measured by applying a sufficiently large current bias and by measuring the voltage drop $V$. (\textit{Inset}) Fabry-Pérot oscillations as a function of $V_{BG}$ over the range highlighted by the rectangle in the main panel. A polynomial background is removed as discussed in the Supporting Information, Section S1.1. ($\mathbf{d}$) \textit{Upper panel} Schematics of the doping across the device in the p-type (hole) doping regime: Dirac cones in red are traced according to the local doping ($\mu$ is the chemical potential). The n-type (electron) doping induced by Nb leads to the formation of a n-p-n cavity. \textit{Lower panel} Sketch of ballistic transport across the device. Charge carriers have a finite probability to be reflected at the p-n interfaces, leading to Fabry-Pérot interference analogous to an optical cavity. ($\mathbf{e}$) Voltage drop $V$ as a function of DC current bias $I_{bias}$ and backgate voltage $V_{BG}$. The current sweep direction is indicated by the black arrow. The dotted blue and red lines correspond respectively to the retrapping and switching currents. $T = 40$~mK, $B=0$~T (\textit{Inset}) $V$-$I_{bias}$ curve for $V_{BG}=40$~V showing switching-retrapping behavior. Arrows indicate the sweep direction. ($\mathbf{f}$) Differential resistance $dV/dI$ (obtained by numerical differentiation of the measured DC voltage drop with respect to the applied DC current bias) as a function of the out-of-plane magnetic field $B$ (in the $\pm10$~mT range) and current bias $I_{bias}$ at $V_{BG}=40$~V, displaying a Fraunhofer interference pattern. Darker points, indicating a peak in $dV/dI$, correspond to the transition from the dissipationless to the dissipative regime. $T = 40$~mK.}
\end{figure}

We experimentally investigate the QH-SC coexistence using a back-gated monolayer graphene Josephson junction, whose structure is schematically depicted in Fig.~\ref{fig1}a. The Josephson junction has length $L=400$~nm and width $W=3\;\mu$m (in line with the device dimensions in Ref.~\citenum{Amet2016}); an optical microscopy image of the device is shown in Fig.~\ref{fig1}b together with a sketch of the measurement configuration. All measurements are performed in a Leiden Cryogenics dilution refrigerator equipped with cryogenic filtering, at a base temperature of $40$~mK (except otherwise indicated). We employ high-mobility graphene single crystals grown by chemical vapor deposition (CVD) \cite{Miseikis_2015, Miseikis_2017,Banszerus2015, Pezzini_2020} and encapsulated in hBN flakes via the dry van der Waals pickup method \cite{Wang2013, Purdie2018, Pezzini_2020}. Top and bottom hBN flakes are each 30~nm thick, resulting in a total stack thickness $t=60$~nm.
High-transparency graphene-Nb interfaces are obtained by combining dry etching of hBN \cite{Wang2013} with DC magnetron sputtering of 60-nm thick Nb contacts \cite{BenShalom2016} (further details on the fabrication process are reported in the Methods section). Nb maintains its superconducting state at fields well exceeding the QH onset for high-mobility graphene (e.g., a QH onset as low as 50~mT was measured in Hall bar devices based on the same graphene crystals in Ref.~\citenum{Pezzini_2020}). The device is fabricated on a \ce{Si/SiO_2} substrate ($285$-nm thick \ce{SiO_2}), and the underlying p-doped Si functions as a back-gate (BG).

When measuring the normal-state resistance of the device as a function of the BG voltage ($V_{BG}$) we obtain asymmetric curves, as shown in Fig.~\ref{fig1}c, with the charge neutrality point (CNP) shifted to slightly negative $V_{BG}$ ($-3.6$~V in this specific sample). This shift is attributed to Fermi level pinning at the Nb contacts, that results in n-type (electron) doping in the neighboring graphene region, as shown in Fig.~\ref{fig1}d. For positive $V_{BG}$ the resistance $R$ approaches the Sharvin limit \cite{BenShalom2016} (dashed black line in Fig.~\ref{fig1}c), indicating high interface transparency (up to $Tr\sim 0.8$ estimated following Ref.~\citenum{BenShalom2016}). For $V_{BG}<V_{CNP}$, a n-p-n cavity forms across the junction \cite{Calado_2015}, leading to a larger resistance on the p-type (hole) doping side. This stems from the p-n interfaces acting as partially reflecting barriers (as exemplified in Fig.~\ref{fig1}d), with an estimated transparency $Tr\sim 0.3$ at $V_{BG}=-40$~V. In this regime, Fabry-Pérot (FP) oscillations are observed (see inset to Fig.~\ref{fig1}c). These oscillations are a clear signature of ballistic transport as they can occur only if electrons maintain coherence while traveling across the device \cite{Young2009, Calado_2015,BenShalom2016}. In ballistic n-p-n cavities, the resistance oscillates as a function of the Fermi wavevector $k_F$, with minima satisfying the resonance condition $k_F L_c=m\pi+\pi/2$ (where $L_c$ is the cavity length and $m$ the cavity mode number) \cite{Calado_2015}. From the periodicity of the oscillations in the p-type doping regime we estimate the cavity length to be $\sim200$~nm. A comparable cavity length value is estimated also in the n-type doping regime, where a n-n'-n cavity forms and FP oscillations are observed. The oscillations are, however, much less visible owing to the higher interface transparency (further details are provided in Supporting Information, Section S1.1).

Figure~\ref{fig1}e shows a colormap of the voltage drop $V$ across the junction, measured while sweeping the DC current bias ($I_{bias}$) at different $V_{BG}$ values in the $(-40,40)$~V range. The DC bias sweep direction is indicated by the black arrow. Supercurrent is measured in the white central region corresponding to zero voltage drop. The critical current ($I_c$, identified as the boundary between the dissipationless and dissipative regions) is modulated by $V_{BG}$ and approaches values up to 10 $\mu$A at large n-type doping (red star in Fig.~\ref{fig1}e).

This value and the corresponding supercurrent density $J_s\simeq 3.3\;\mu$A$\mu$m$^{-1}$ are in line with graphene Josephson junctions with similar length, as shown in Refs.~\citenum{BenShalom2016,Amet2016}, demonstrating comparable sample quality. A minimum of $I_c$ is observed at the CNP, while on the p-doping side $I_c$ is consistently smaller than on the corresponding n-doping side, due to the reduced transparency at the p-n interfaces. FP oscillations are visible also in the supercurrent as a modulation of $I_c$: in this case, maxima in the supercurrent correspond to minima in the resistance (see details in Supporting Information, Section S1.2). The device exhibits a pronounced switching-retrapping behavior (as can be seen also from the $V$-$I_{bias}$ curve in the inset in Fig.~\ref{fig1}e). This behavior is known to originate from electron heating in the dissipative branch \cite{Courtois2008,Borzenets2013,Borzenets2016}. Additional characterization of the device at zero magnetic field, which includes a discussion about multiple Andreev reflections and the $I_c\times R_n$ figure of merit is reported in Section S2 of the Supporting Information.

In the low magnetic field limit, for $|B|<10$~mT, a Fraunhofer interference pattern (Fig.~\ref{fig1}f) is observed. Deviations from a perfectly regular pattern can be attributed to slight inhomogeneities in the supercurrent distribution in the direction perpendicular to the current flow \cite{Dynes_Fulton1971}. Additionally, in the Meissner phase at low magnetic fields, flux focusing is known to increase the effective flux through the junction \cite{Paajaste2015}, leading to a smaller periodicity than expected \cite{BenShalom2016,tinkham_1996, Meissner1933}. Following the argument in Ref.~\citenum{Amet2016}, based on the geometry of our device we expect a focusing factor $\simeq 1.8$ (defined as the ratio between the effective field in the junction and the applied field $B$). This value is in good agreement with $\simeq 1.7$, estimated as the ratio between the expected theoretical position of the first Fraunhofer minima and their observed position (for further details see the Supporting Information, Section S3).

As the magnetic field increases, the amplitude of the SC does not follow the expected suppression based on the standard Fraunhofer dependence. Instead, regions with supercurrent persist. These regions, known as "superconducting pockets" \cite{BenShalom2016}, are characterized by a partial or complete suppression of the differential resistance and can survive up to large magnetic fields reaching the Tesla range. In the semiclassical regime, below the onset of QH states, where ${r_c>L/2}$ (${r_c=\hbar k_F/eB}$ is the cyclotron radius), we observe superconducting pockets not only in the n-type doping regime, as reported in Ref.~\citenum{BenShalom2016}, but also for p-type doping. Additional discussion regarding the SC pockets in the semiclassical regime is reported in the Supporting Information, Section S4. SC pockets are observed in the QH regime, as well, as discussed in the following.

\subsection*{Supercurrent up to 2.4~T}

\begin{figure}[H]
	\centering
	\includegraphics[width=\linewidth]{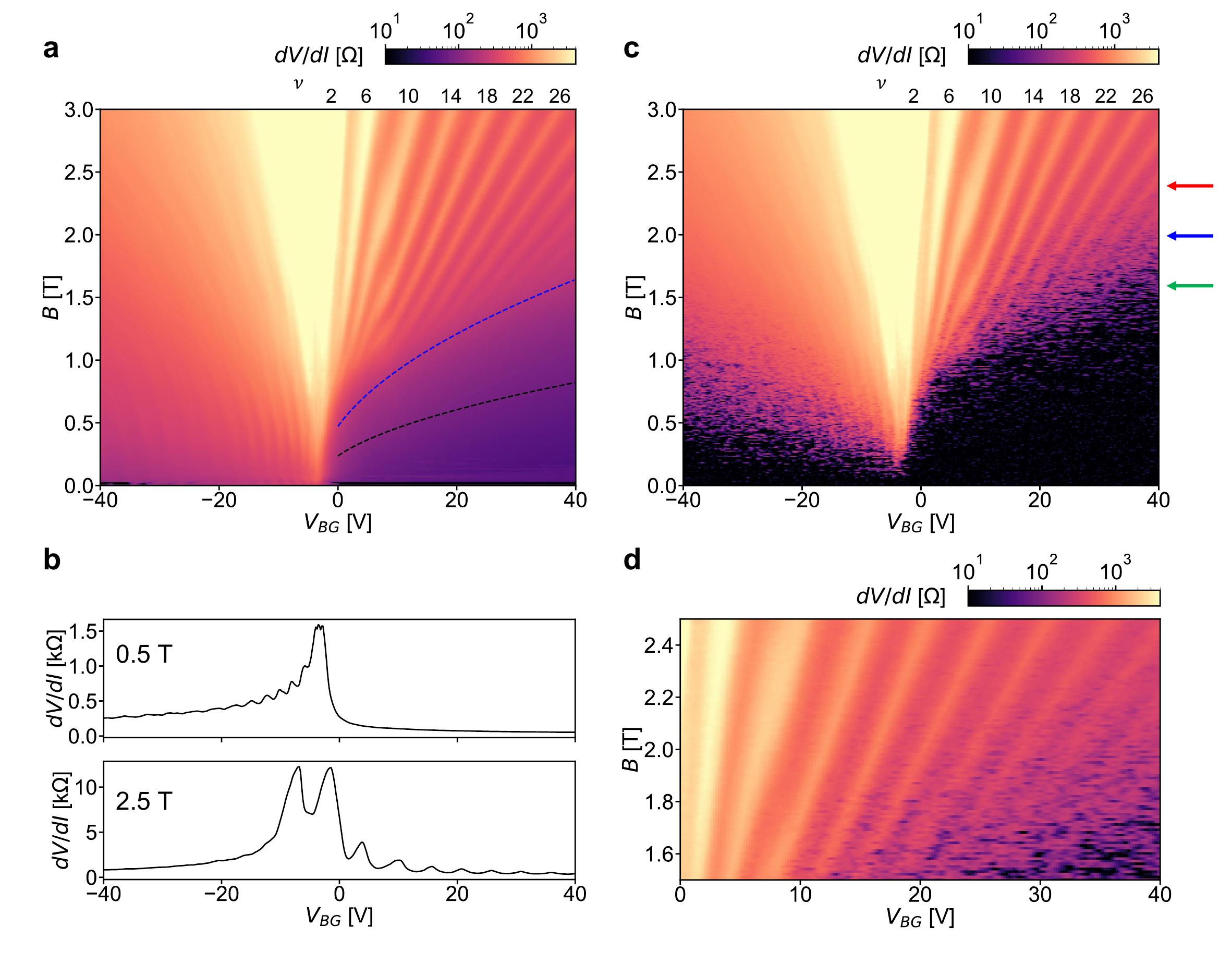}
	\caption{\label{fig2} ($\mathbf{a}$) Landau fan diagram ($dV/dI$ as a function of $B$ and $V_{BG}$). The differential resistance $dV/dI$ is measured with a lock-in amplifier, by applying a $100$~nA AC bias excitation on top of a fixed 200~nA DC bias excitation. The dashed black line represents the expected QH-semiclassical regime crossover calculated from the geometrical dimension $L=400$~nm of the device, while the blue dashed line is obtained by considering the estimated FP cavity length $L_c=200$~nm, as explained in the main text. The two lines are obtained from the equation $\hbar k_F/eB=L/2$, respectively for $L=400$~nm and $L=200$~nm. The Fermi wavevector $k_F=\sqrt{\pi n}$ is calculated from the gate-dependent carrier density $n=f( V_{BG}-V_{CNP})$ ($f$ is the gate lever arm). ($\mathbf{b}$) Linecuts from panel \textbf{a} of $dV/dI$ vs.~$V_{BG}$ at fixed magnetic field. At $B=0.5$~T, on the p-type doping side, FP oscillations are still observed, as they evolve with magnetic field. At $B=2.5$~T, $dV/dI$ shows a non-monotonic behaviour, as discussed in text. ($\mathbf{c}$) Landau fan diagram obtained by measuring the differential resistance $dV/dI$ with a lock-in amplifier by applying a small (500~pA) AC bias and no DC bias. QH superconducting pockets appear as dark spots on top of $dV/dI$ oscillations, absent in panel $\mathbf{a}$. $T = 40$~mK. The coloured arrows indicate the position of the acquisitions shown in Fig.~\ref{fig3}. ($\mathbf{d}$) Zoom of the region in \textbf{b} where SC pockets are seen in the QH regime, appearing as darker spots indicating locally suppressed $dV/dI$.}
\end{figure}

The plot in Fig.~\ref{fig2}a shows the differential resistance $dV/dI$, measured with a lock-in amplifier as a function of BG voltage $V_{BG}$ and magnetic field $B$, while applying a relatively large AC current of $100$~nA on top of a $200$~nA DC bias. Linecuts of $dV/dI$ at fixed $B=0.5$~T and $B=2.5$~T are reported in Fig.~\ref{fig2}b. On the n-type doping side, to the right of the CNP, a regular Landau fan diagram spreads out from the CNP, while the pattern on the p-type side is heavily influenced by the p-n interfaces discussed above, with FP oscillations merging with Landau levels, and following a $B^2$ dispersion with respect to $k_F$, as discussed in the Supporting Information, Section S1.2. The observation of FP oscillations at finite magnetic field provides further evidence of the ballistic nature of electrical transport. Given the wide aspect ratio of the junction ($L/W=7.5$), $dV/dI$ shows an oscillating behavior with maxima corresponding to the position of QH plateaus \cite{Abanin2008, Williams2009} and reduced differential resistance within them (as can be also seen in the linecut at $B=2.5$~T shown in Fig.~\ref{fig2}b).

At large magnetic field values, in the n-type doping regime ($V_{BG}>V_{CNP}$), the condition $r_c= L/2$ marks the crossover between the semiclassical and the QH regime. In our device, however, the boundary appears to be determined by a length scale shorter than the junction length, which we identify as the FP cavity length. This is shown in Fig.~\ref{fig2}a where the separation line corresponding to the junction geometrical length, $L=400$~nm (black dashed line) clearly does not match the semiclassical to QH regime separation. Instead, considering the FP cavity length (blue dashed line, corresponding to $200$~nm), a better agreement is obtained.

In the data shown in Fig.~\ref{fig2}a the applied current bias (comprising a $100$~nA AC excitation on top of a DC $200$~nA excitation) is large enough to suppress any trace of supercurrent. By lowering the AC bias down to 500~pA and setting the DC bias to zero, numerous superconducting states are observed, not only in the semiclassical regime, but also in the QH regime, as shown in Fig.~\ref{fig2}c. In Fig.~\ref{fig2}c the SC pockets can be identified as dark spots on top of the Landau fan diagram, absent in the large bias acquisition of Fig.~\ref{fig2}a. They correspond to a partial local suppression of the differential resistance and can be appreciated by directly comparing Fig.~\ref{fig2}c with Fig.~\ref{fig2}a. A zoom of the region of the Landau fan diagram where SC pockets are observed is reported in Fig.~\ref{fig2}d.

\begin{figure}[!htb]
	\centering
	\includegraphics[width=0.5\linewidth]{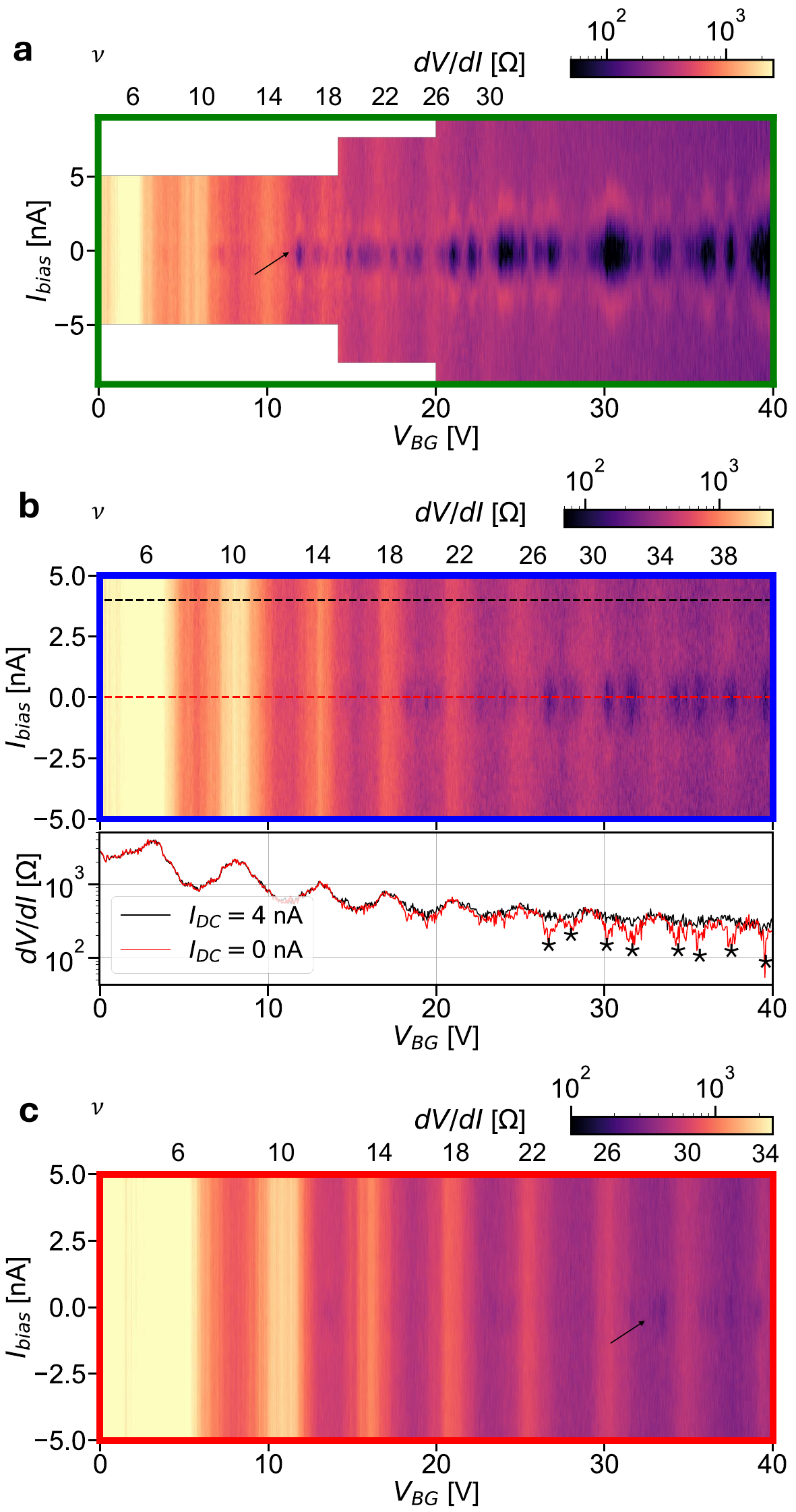}
	\caption{\label{fig3} Differential resistance $dV/dI$ as a function of backgate voltage $V_{BG}$ and applied DC bias $I_{bias}$ for selected values of the magnetic field $B$. $dV/dI$ was measured with a lock-in amplifier by applying a small AC bias excitation, as indicated in the following. ($\mathbf{a}$) Acquisition at $B=1.6$~T, corresponding to the green arrow in Fig.~\ref{fig2}c. AC modulation: 100 pA in the $(0, 14)$ V range, 150 pA in the $(14,20)$ V range, 200 pA in the $(20, 40)$ V range. Black arrow indicates the chosen pocket for the acquisitions shown in Fig.~\ref{fig4}. ($\mathbf{b}$) Acquisition at $B=2.0$~T, corresponding to the blue arrow in Fig.~\ref{fig2}c. AC bias: 100 pA. Linecuts at ${I_{DC}=0}$~nA (red line) and ${I_{DC}=4}$~nA (black line), are shown in the lower panel. Superconducting pockets, indicated by *, appear as a partial suppression of $dV/dI$ in the zero DC bias line compared to the finite DC bias case. ($\mathbf{c}$) Acquisition at $B=2.4$~T, corresponding to the red arrow in Fig.~\ref{fig2}c. AC bias: 100 pA. A weak suppression of $dV/dI$ at large filling factor can be seen (indicated by the black arrow).}
\end{figure}

To further substantiate these observations of QH-SC, we performed measurements of $dV/dI$ with AC excitation of 100-200~pA, as a function of $V_{BG}$ and $I_{bias}$, for selected values of magnetic field. These acquisitions are reported in Fig.~\ref{fig3}. Fig.~\ref{fig3}a shows data for $B=1.6$~T (corresponding to the green arrow in Fig.~\ref{fig2}c). Some pockets are observed in the QH regime, starting at ${\nu>14}$, with $I_c$ in the $1$~nA range (similar to Refs.~\citenum{Amet2016,Vignaud2023}). For $V_{BG}\gtrsim 25-30$~V, which brings the system close to the QH onset, pockets with larger $I_c$ are also observed. At $B=2$~T (Fig.~\ref{fig3}b, blue arrow in Fig.~\ref{fig2}c) multiple pockets are observed at relatively large filling factor. These pockets are mainly located between $dV/dI$ peaks, i.e., in the transition region between contiguous QH plateaus. The lower panel in Fig.~\ref{fig3}b compares the line cuts taken at $I_{bias}=0$ nA and $I_{bias}=4$ nA (the latter large enough to suppress the supercurrent). Several SC states, marked with "*", can be seen as suppression of $dV/dI$ in the red line as compared to the black one. At $B=2.4$~T (Fig.~\ref{fig3}c, red arrow in Fig.~\ref{fig2}c), residual supercurrent regions are present for $\nu>26$ (see arrow in Fig.~\ref{fig3}c). The amplitude of the pockets is rapidly suppressed with increasing temperature, and no supercurrent is observed for $T > 200$~mK (see Supporting Information, Section S5 for details).

\subsection*{Periodicity of supercurrent at QH transitions}

Previous reports of supercurrent in the QH regime showed a clear periodicity of the SC pockets with magnetic field. For junctions with a large aspect ratio similar to our device \cite{Amet2016,Draelos2018,Seredinski2019}, a $\Phi_0=h/2e$ periodicity was observed as a function of magnetic field, for fixed backgate voltage. In those cases, the SC was located on top of QH plateaus, hence it was carried exclusively by channels along the sample edges. In our experiment, instead, the suppression of $dV/dI$ is not observed on the QH plateaus but in the transition regions between the plateaus. Although signatures of a SC at QH transitions are visible in data presented in Refs.~\citenum{Draelos2018,Vignaud2023}, a specific investigation of this phenomenon is currently lacking.

\begin{figure}[H] 
	\centering
	\includegraphics[width=\linewidth]{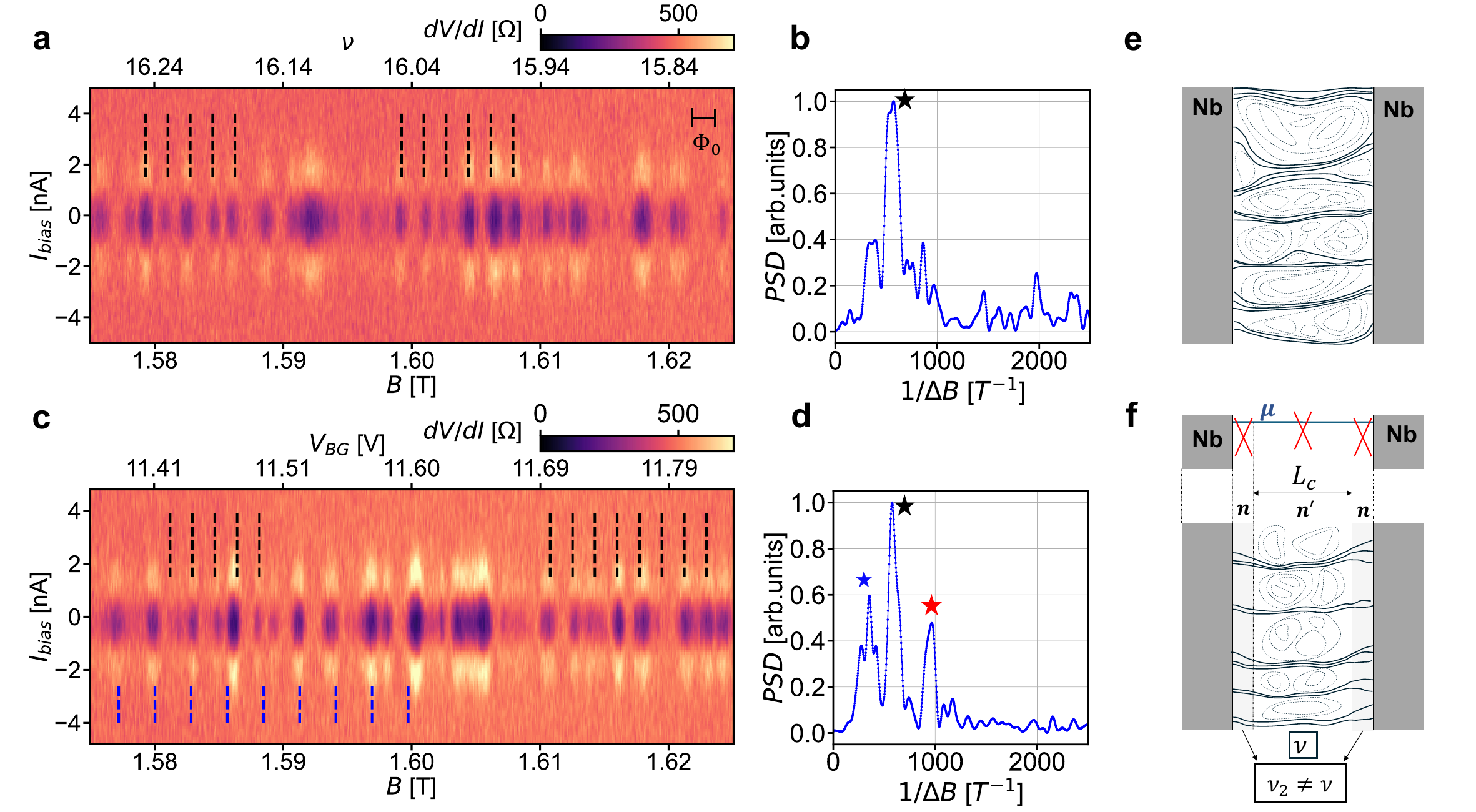}
	\caption{\label{fig4} ($\mathbf{a}$) Differential resistance $dV/dI$ as a function of magnetic field $B$ and DC bias current $I_{bias}$, acquired at fixed $V_{BG}=11.6$~V. $dV/dI$ was measured with a lock-in amplifier by applying a 100 pA AC bias excitation on top of the sweeping DC bias. Black dashed lines correspond to $\Phi_0$-periodic oscillations on the full junction's area (main peak in the Fourier spectrum in panel $\mathbf{b}$). ($\mathbf{b}$) Fourier transform (power spectral density) of zero-bias $dV/dI$ data in panel \textbf{a}. ($\mathbf{c}$) Same as panel $\mathbf{a}$, acquired at fixed filling factor $\nu=16.0$. Black dashed lines correspond to $\Phi_0$-periodic oscillations on the full junction's area (main peak in the Fourier spectrum in panel $\mathbf{d}$, indicated by the black star), while blue dashed lines correspond to $\Phi_0$-periodic oscillations on the smaller area corresponding to the FP cavity (peak in the Fourier spectrum in panel $\mathbf{d}$ indicated by the blue star). ($\mathbf{d}$) Same as panel $\mathbf{b}$, for dataset in \textbf{c}. ($\mathbf{e}$) Schematics of the transport mechanism in the percolation regime: random transport channels arise in the bulk via merging of QH droplets. ($\mathbf{f}$) Same as panel $\mathbf{e}$, considering the highly n-doped regions near the Nb contacts.}
\end{figure}

To elucidate these differences, we performed two dedicated measurements, shown in Fig.~\ref{fig4}: one at fixed $V_{BG}$ (Fig.~\ref{fig4}a) and one at fixed filling factor $\nu$ (Fig.~\ref{fig4}c). The acquisitions are performed around $B=1.6$~T in a 50 mT-wide magnetic field range, around the pocket indicated by the arrow in Fig.~\ref{fig3}a. In the first acquisition, shown in Fig.~\ref{fig4}a, $V_{BG}=11.6$~V is kept constant and the magnetic field is varied, causing a change of the filling factor from $\nu=15.8$ to $\nu=16.3$. In the fixed-$\nu$ acquisition, shown in Fig.~\ref{fig4}c, magnetic field and backgate voltage are varied simultaneously to keep $\nu=16.04\pm0.02$ (in the transition region between the plateaus at $\nu=14.0$ and $\nu=18.0$). No sharp periodicities emerge in both Fig.~\ref{fig4}a and \ref{fig4}c. However, an approximate periodicity can be appreciated from bunches of pockets that appear evenly spaced (see the guiding dashed lines).

To further investigate these observations we performed a Fourier analysis on the $dV/dI$ signal at zero DC bias. The resulting Fourier spectra are shown in Fig.~\ref{fig4}b (relative to Fig.~\ref{fig4}a) and Fig.~\ref{fig4}d (relative to Fig.~\ref{fig4}c). In the fixed-$V_{BG}$ acquisition of Fig.~\ref{fig4}a, a main peak (black star in Fig.~\ref{fig4}b) appears at $1/\Delta B=575$~T$^{-1}$: this value corresponds to a period of ${\Delta B\simeq 1.7}$~mT, which in turn corresponds to a flux variation of ${\Delta\Phi=\Delta B \times L\times W\simeq\Phi_0}$, where $L=400$~nm and $W=3\;\mu$m are the physical dimensions of the device (note that, for ${B>B_{c,1}\simeq 180}$~mT \cite{Uday2024}, magnetic field penetrates the contacts, hence flux focusing does not play a role anymore). Dashed black lines in Fig.~\ref{fig4}a are traced according to this periodicity. 
We observe that superconducting pockets cluster into approximately equally-spaced groups, interspersed with regions where their distribution appears more random. In the following we propose a mechanism which could explain the observed quasi-periodicity.

As previously mentioned, the observed superconducting pockets are located in the transition region between QH plateaus, where electrical currents percolate across merging QH droplets, as exemplified in Fig.~\ref{fig4}e. Near zero magnetic field, where interference patterns such as the one in Fig.~\ref{fig1}f are measured, multiple parallel transport channels are available throughout the entire device. The magnetic flux through the junction determines the interference between these channels \cite{tinkham_1996}, leading to the observation of maxima and minima as a function of magnetic field (i.e., the Fraunhofer pattern). In the percolative QH regime, exemplified in Fig.~\ref{fig4}e, the supercurrent density can spatially arrange in a way similar to the near-zero magnetic field state, i.e., it can distribute across the whole 2D bulk, though non-uniformly. As a result, superconducting interference as a function of the magnetic field leads to the observed $\Phi_0$-periodicity reported in Fig.~\ref{fig4}a,b. Since the shape and distribution of QH droplets evolves as the filling factor is varied (Fig.~\ref{fig4}a and \ref{fig4}b), the distribution of transport channels also changes, as discussed above. This evolution is driven by the details of the potential fluctuations across the graphene sheet, which likely account for the observed irregular quasi-periodicity. The available transport channels, forming randomly in the percolation regime within the bulk of the junction, are dephased with respect to each other. Therefore, one expects the supercurrent to be of the order of that of a single channel $I\sim ev_F/L$. Indeed, all measured pockets in the QH regime show a characteristic critical current of $\sim 1$~nA, consistent with previous reports of QH-SC \cite{Amet2016,Vignaud2023}, and in good agreement with the expected value $\sim ev_F/L$, taking into account the reduced Fermi velocity of QH channels \cite{Ronen2021,Deprez2021} in proximity to superconductors \cite{Vignaud2023}.

In principle, the re-arrangement of the QH droplets should be prevented by keeping the filling factor value fixed, but no sharp $\Phi_0$-periodicity is observed even in the constant-$\nu$ data (Fig.~\ref{fig4}c-\ref{fig4}d). On the contrary, as for the fixed-$V_{BG}$ acquisition of Fig.~\ref{fig4}a,b, a main peak in the Fourier transform appears (black star in Fig.~\ref{fig4}d), corresponding to an approximate $\Phi_0$-periodicity. A second smaller peak develops at lower frequency (blue star in Fig.~\ref{fig4}d). We interpret this peak as evidence of $\Phi_0$-periodic oscillations stemming from a smaller area, whose dimensions correspond to the FP cavity previously discussed. As pointed out earlier, a n-n'-n cavity forms in the n-type doping regime, as sketched in Fig.~\ref{fig4}f. In the central area of the junction ($n^\prime$) the filling factor value is kept constant by acting on $B$ and $V_{BG}$ in the constant-$\nu$ measurement configuration. However, the areas near the contacts host a different charge density that is determined by the interplay of Fermi-level pinning at Nb contacts and applied $V_{BG}$. While at large magnetic fields FP interference is no longer observed due to cyclotron motion, the doping variation across the sample is preserved as it is determined by the effect of the Nb contacts. As a result, the local filling factor in the region close to the Nb contacts ($\nu_2$) varies as the magnetic field and $V_{BG}$ are changed. Consequently, the QH droplets continue to evolve in these regions, leading to an uncontrolled re-arrangement of the available transport channels. The extracted frequency for the second peak in Fig.~\ref{fig4}d, identified by the blue star and centered at $355$~T$^{-1}$, corresponds to an area of width $W=3\;\mu$m and length of $245\pm 30$~nm. Dashed blue lines in Fig.~\ref{fig4}c are traced according to this periodicity. This length value is comparable to the estimated FP cavity length ($\sim 200$~nm), which also appears to determine the semiclassical-to-QH regime separation.
Altogether, these observations indicate an interplay between the cavity originated by doping variations and the spatial distribution of SC-carrying percolative paths. A third peak (red star in Fig.~\ref{fig4}d) is also observed in the fixed-$\nu$ acquisition, located at a frequency value ascribable to the sum of the main oscillatory components. Such peak arises due to a coupling of the two identified periodicities, implying a net modulation dominated by the smaller supercurrent (further details are discussed in the Supporting Information, Section S6).

Percolative superconductivity is a phenomenon that is observed in various solid-state systems. It plays a key role in the behavior of quantum materials such as cuprates \cite{Pelc2018} and oxide interfaces \cite{Caprara2013}, but also granular superconductors \cite{Strelniker2007} of high technological relevance \cite{Grunhaupt2019}. In our experiment, we realize a synthetic percolative superconductor obtained via Josephson effect at QH transitions, which could serve as testbed for studying this transport regime, with the added flexibility provided by van der Waals devices. For example, we note that the scaling behavior of QH transitions in graphene appears to depend on the employed substrate (such as SiO$_2$ \cite{Amado_2012}, hBN \cite{Cobaleda2014}, hBN-on-graphite \cite{Kaur2024}). This suggests that substrate engineering could offer a means of tuning the induced SC.

\section*{Conclusions and outlook}

In conclusion, we have demonstrated induction of superconducting states in the QH regime in Nb-contacted graphene Josephson junctions. We observe quasi-$B$-periodic SC oscillations, with a main peak in the Fourier spectrum at $\Phi_0$. Since superconducting pockets are present in the QH percolative regime, our findings point to an interference mechanism analogous to the low-field Fraunhofer pattern. Additionally, a second peak at lower frequency is detected when the filling factor in the central area of the sample is kept constant, indicating that interference takes place also over a smaller area, corresponding to the FP cavity, which is influenced by local doping at the Nb-graphene interfaces. Our findings provide evidence for a distinct mechanism for SC-QH coexistence, differing from edge-mediated transport \cite{Amet2016} and $2\Phi_0$-periodic chiral supercurrent \cite{Vignaud2023}, with possible general implications for the investigation of percolative superconductors.

\section*{Methods: device fabrication}
We assemble the hBN-graphene-hBN stack using standard dry pick-up methods \cite{Wang2013,Purdie2018}. We employ a poly(bisphenol A carbonate) (PC) film deposited onto polydimethylsiloxane (PDMS) \cite{Purdie2018} to combine micromechanically exfoliated hBN flakes and CVD-grown graphene as described in Ref.~\citenum{Pezzini_2020}. After the assembly, we select target areas for device fabrication based on Raman spectroscopy signatures indicating minimal nanoscale variations \cite{Couto2014,Neumann2015,Pezzini_2020}. A polymethylmethacrylate (PMMA) mask is used for electron beam litography (EBL) patterning. A first EBL step defines self-aligned one-dimensional superconducting contacts \cite{BenShalom2016}. A $15$~s mild oxygen plasma step (10 W power) is performed prior to etching to ensure complete removal of polymer residues in the exposed areas. A mixture of \ce{CF_4} and \ce{O_2} (flowrate 20 sccm and 2 sccm respectively) is used to etch through the entire 60 nm-thick stack in approximately 30 s at 25 W power. An additional 10 s mild oxygen plasma step is performed after \ce{CF_4}-\ce{O_2}, yielding an increased Nb-graphene interface transparency (by typically $20-30$~\%). 60 nm-thick Nb contacts are then deposited using DC magnetron sputtering at a rate of $\sim 1$~nm/s, with no adhesion layer deposited prior to Nb. Lift-off is performed in warm acetone ($T=50^\circ$C) for 20 minutes. A second EBL patterning is used to define the device mesa, followed by the same RIE process used for the contacts. The sample is eventually cleaned in acetone at room temperature overnight.

\section*{Associated content}
\subsection*{Supporting information}

Additional device characterization measurements: Fabry-Pérot oscillations (Figs.~S1, S2, S3, S4), multiple Andreev reflection and $I_c\times R_n$ product (Fig.~S5), flux focusing (Fig.~S6); superconducting pockets in the semiclassical regime (Fig.~S7); temperature dependence of pocket amplitude in the QH regime (Figs.~S8,S9); coupling of B-modulated supercurrents and sum frequency in the FFT spectrum (Figs.~S10, S11, S12, S13).

\section*{ACKNOWLEDGEMENTS}
We acknowledge financial support from the PNRR MUR Project PE0000023-NQSTI funded by the European Union-NextGenerationEU. F.G. acknowledges the EU’s Horizon 2020 Research and
Innovation Framework Programme under Grants No. 964398 (SUPERGATE) and No. 101057977 (SPECTRUM) for partial financial support. K.W. and T.T. acknowledge support from the JSPS KAKENHI (Grant Numbers 21H05233 and 23H02052) , the CREST (JPMJCR24A5), JST and World Premier International Research Center Initiative (WPI), MEXT, Japan.

\newpage

\begin{center}
	\section*{Supporting information}
\end{center}

\beginsupplement

\section{Fabry-Pérot oscillations}

\subsection{Estimating the FP cavity length from FP oscillations}

\begin{figure}[!htb]
	\centering
	\includegraphics[width=\linewidth]{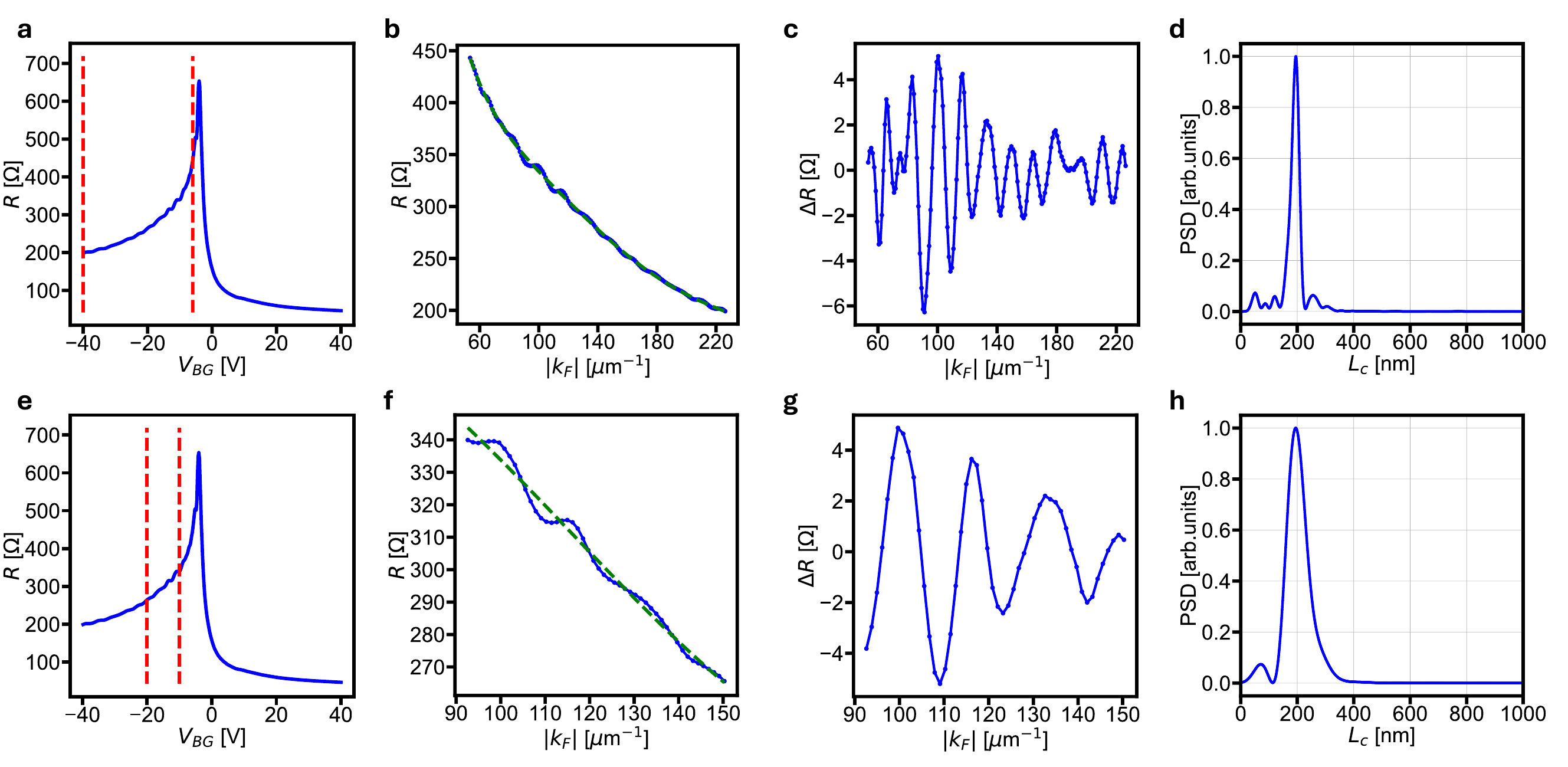}
	\caption{Analysis on FP oscillations in the p-type doping regime. ($\mathbf{a}$) Normal state resistance as a function of $V_{BG}$. Vertical red dashed lines indicate the selected interval. ($\mathbf{b}$) Normal state resistance $R$ as funtion of Fermi wavevector $k_F$ in the selected interval. The fitted polynomial background is shown as green dashed line. ($\mathbf{c}$) $R$ \textit{versus} $k_F$ signal, detrended of the background (here and in the main text labelled as $\Delta R$). ($\mathbf{d}$) Fourier transform (power spectral density) of data in panel \textbf{c}. Peak corresponds to $L_c\sim 200$~nm. ($\mathbf{e}$-$\mathbf{h}$) Same as $\mathbf{a}$-$\mathbf{d}$, but in a $10$~V wide window ($V_{BG}=(-10,-20)$~V). \label{fig_FP_pdop}}
\end{figure}

In ballistic Josephson junctions, FP oscillations are observed as modulation of the resistance. Given the resonance condition $k_F L_c=m\pi+\pi/2$, the cavity length can be estimated from the period of resistance oscillations as a function of $k_F$, obtained by means of a Fourier transform. An example is shown in Fig.~\ref{fig_FP_pdop}, where we report the FP cavity length calculation in the p-type doping regime. The normal state resistance $R$ values are plotted \textit{versus} Fermi wavevector $k_F$ in \textbf{b}. The Fermi wavevector $k_F=\sqrt{\pi n}$ is calculated from the gate-dependent carrier density $n=f(V_{BG}-V_{CNP})$ ($f$ is the gate lever arm). Due to the requirement of an evenly spaced x-array to apply the Fourier transform, an interpolation procedure is applied to redefine the $R$ \textit{versus} $k_F$ dataset such that $k_F$ array is evenly spaced. To isolate FP oscillations from the background, a polynomial function (green dashed line in \textbf{b}) is fitted and then subtracted: resulting data are shown in Fig.~\ref{fig_FP_pdop}c (the same is done in the inset of Fig.~1c in the main text). The Fourier transform is computed by using Welch's method (implemented through the \texttt{scipy.signal.welch} Python function). The result is shown in Fig.~\ref{fig_FP_pdop}d: from the peak position we extract a cavity length of $\sim 200$~nm. By performing the same procedure on narrower intervals of $V_{BG}$ we obtain comparable results, as shown in the example for the $V_{BG}=(-10,-20)$~V interval, Fig.~\ref{fig_FP_pdop}e-h (peak corresponds to $\sim 200$~nm).

\begin{figure}[!htb]
	\centering
	\includegraphics[width=\linewidth]{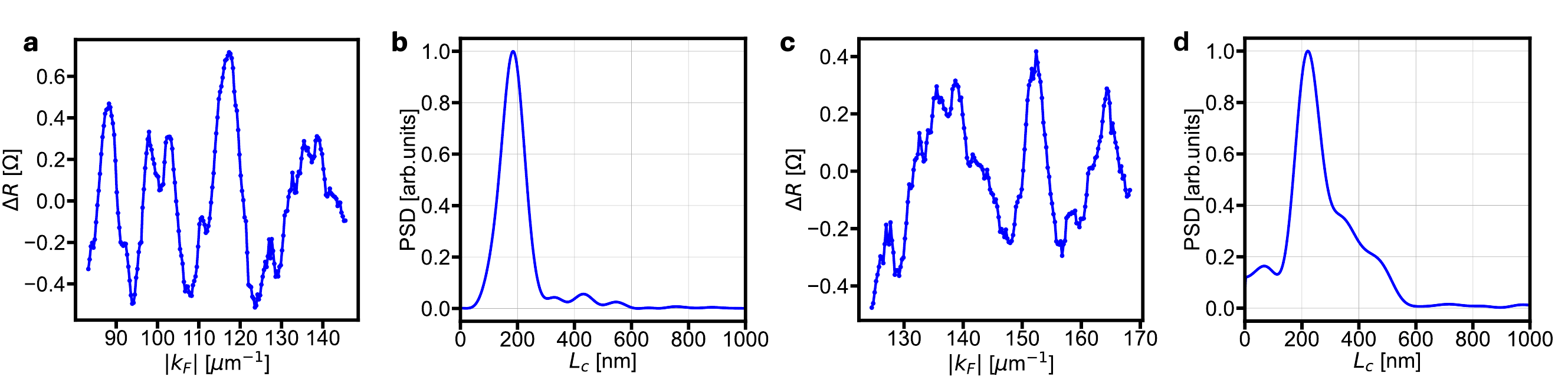}
	\caption{($\mathbf{a}$) Detrended FP oscillations in the $V_{BG}=(0,10)$~V range, plotted as a function of Fermi wavevector. ($\mathbf{b}$) Fourier transform (power spectral density) of data in panel \textbf{a}. Peak corresponds to $\sim 180$~nm. ($\mathbf{c}$) Detrended FP oscillations in the $V_{BG}=(5,15)$~V range, plotted as a function of Fermi wavevector. ($\mathbf{d}$) Fourier transform (power spectral density) of data in panel \textbf{c}. Peak corresponds to $\sim 220$~nm. \label{fig_FP_ndop}}
\end{figure}

In the n-type doping regime FP oscillations, originating in the n-n'-n cavity, have a much lower visibility due to the larger interface transparency (this limits their observability to a few $V$-wide window to the right of the CNP). FP oscillations in the $(0,10)$~V and $(5,15)$~V intervals are shown in Figs.~\ref{fig_FP_ndop}a,c (plotted \textit{versus} $k_F$), the corresponding power spectral densities in Figs.~\ref{fig_FP_ndop}b,d. Peak in \textbf{b} corresponds to $L_c\sim 180$~nm while peak in \textbf{d} to $L_c\sim 220$~nm. Therefore, a cavity length of the order of $\sim 200$~nm is estimated also in the n-type doping regime.

\subsection{FP oscillations in the supercurrent and dispersion with magnetic field}

\begin{figure}[!htb]
	\centering
	\includegraphics[width=\linewidth]{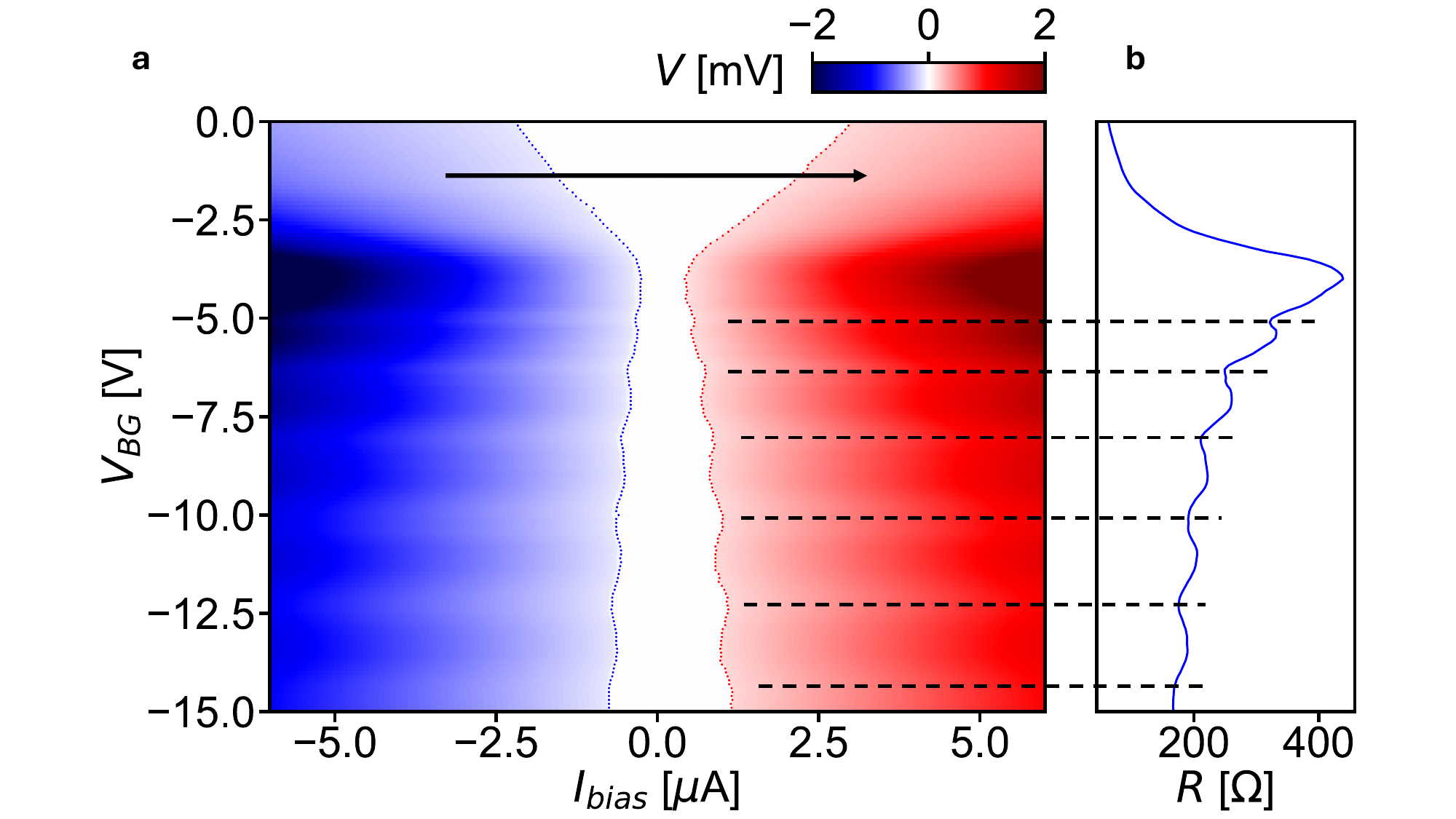}
	\caption{($\mathbf{a}$) Voltage drop as a function of DC current bias $I_{bias}$ and backgate voltage $V_{BG}$ in the p-type doping regime, in the region close to the CNP. Bias sweep direction is indicated by the black arrow. Dotted blue and red lines represent respectively the retrapping and switching currents. ($\mathbf{b}$) Resistance measured at $I_{bias}=-6\;\mu$A. Maxima in the supercurrent correspond to local minima in the resistance, as highlighted by the black guiding dashed lines. ($\mathbf{c}$) \label{fig_FP_SCmap}}
\end{figure}

At zero magnetic field, FP oscillations are observed not only as modulation of the normal state resistance, but also as modulation of the switching and retrapping currents \cite{Calado_2015}. This is shown in Fig.~\ref{fig_FP_SCmap}, where the voltage drop across the junction is plotted as a function of the sweeping DC bias at different $V_{BG}$ values in the interval $(0,-15)$ V. The resistance outside of the supercurrent branch ($I_{bias}=-6\;\mu$A) is shown in the right plot. Local maxima in the supercurrent are aligned to minima in the resistance, as highlighted by the dashed guiding black lines.

\begin{figure}[!htb]
	\centering
	\includegraphics[width=0.6\linewidth]{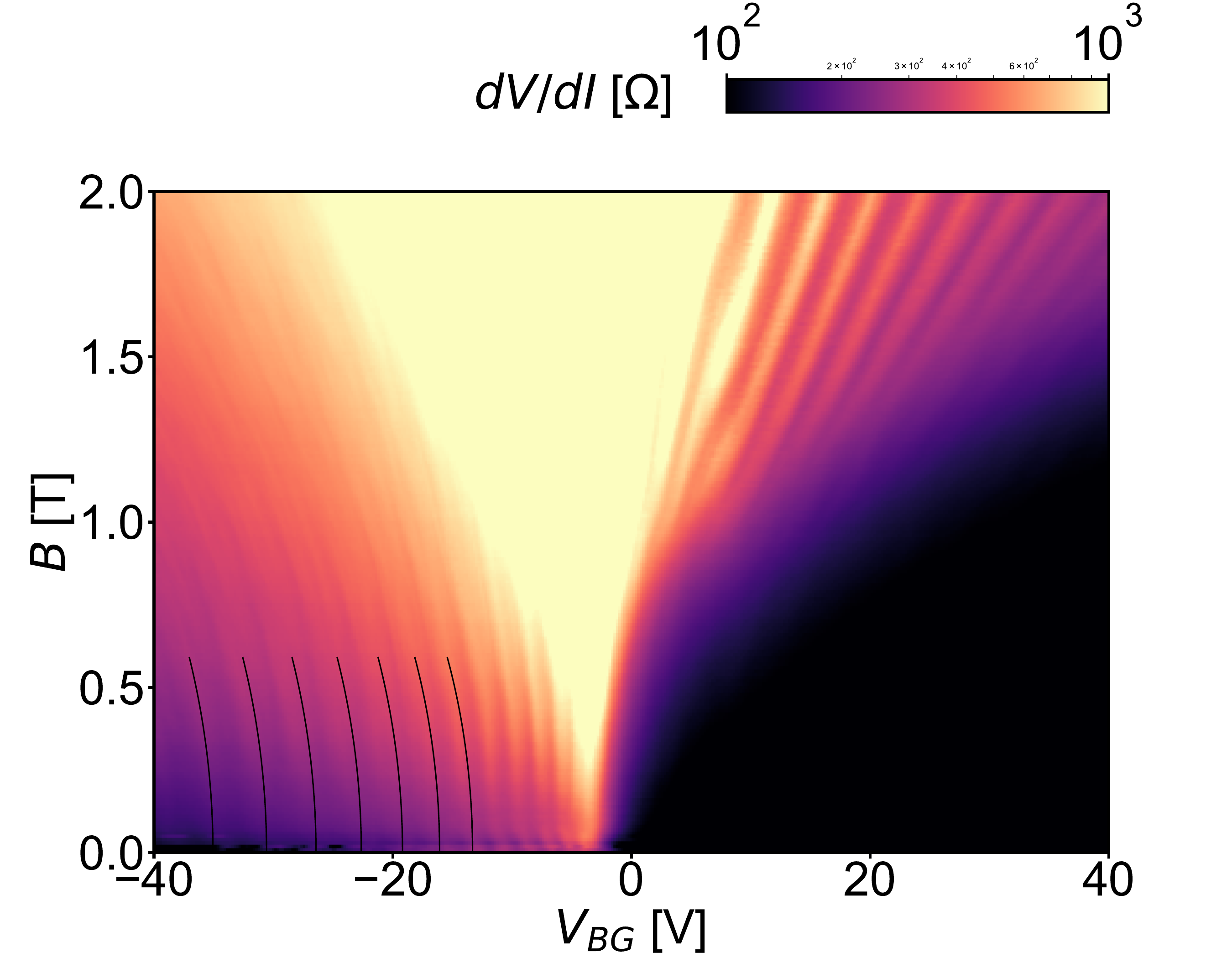}
	\caption{Landau fan diagram as in Fig.2a in main text: the differential resistance $dV/dI$ is measured with a lock-in amplifier by applying a 100 nA AC bias on top of a fixed 200 nA DC bias. Black lines correspond to curves following equation \eqref{eq_FP_B} for cavity mode numbers $m=7-13$. \label{fig_QH_FPdisp}}
\end{figure}

In addition, as discussed for example in Ref.~\citenum{Calado_2015}, in ballistic devices FP oscillations are observed also at finite magnetic field, showing a $B^2$ dispersion with respect to the Fermi wavevector $k_F$. By applying an out-of-plane magnetic field, which results in a Lorentz force to electrons and holes, a shift towards higher charge density of the FP resonances is observed, due to the extra phase accumulation, which depends on the applied field. By considering the phase accumulation in the p-region, the authors in Ref.~\citenum{Calado_2015} obtained a semiclassical expression which links the carrier density (and thus the Fermi wavector $k_F$), the cavity mode number $m$ and the square of the magnetic field $B^2$:

\begin{equation}
	k_FL=m\pi+\frac{\pi}{2}+\frac{\pi}{6m}(L^2\frac{e}{h}B)^2
	\label{eq_FP_B}
\end{equation}	

Dispersion curves corresponding to modes $m=7-13$ are shown in the Landau fan diagram in Fig.~\ref{fig_QH_FPdisp} and show a very good qualitative agreement with the experimental data.

\section{Additional device characterization}
\begin{figure}[!htb]
	\centering
	\includegraphics[width=0.8\linewidth]{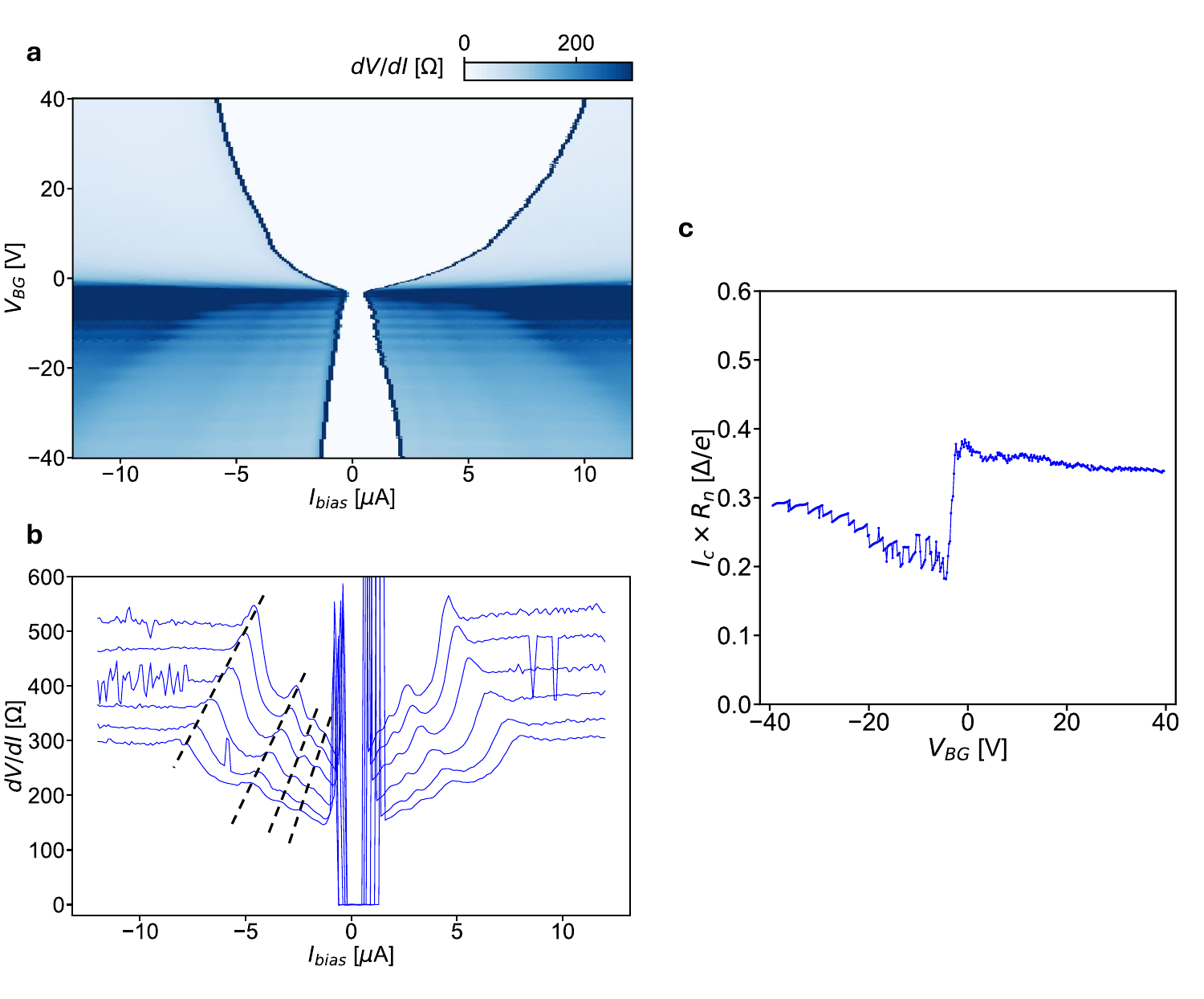}
	
	\caption{($\mathbf{a}$) Differential resistance $dV/dI$ map as function of DC bias $I_{bias}$ and BG voltage $V_{BG}$. $dV/dI$ is obtained by numerical differentiation of data shown in Fig.~1e of the main text. ($\mathbf{b}$) Selected $dV/dI$ curves corresponding to $V_{BG}$ in a $1$~V interval around the CNP point. MAR peaks, here highlighted by the dashed black lines for $n=1-4$, are also visible in the colormap \textbf{a}. ($\mathbf{c}$) $I_c\times R_n$ product, in units of $\Delta/e$. The critical current $I_c$ is obtained from the supercurrent map shown in Fig.~1e, while the normal state resistance is obtained from Fig.~1c of the main text. \label{fig_add_char}}
\end{figure}

Fig.~\ref{fig_add_char}a shows the differential resistance obtained by numerical differentiation of data presented in Fig.~1e of the main text. The supercurrent branch can be identified as the white area at the center of the map corresponding to $dV/dI=0$.

In a Josephson junction, the differential resistance exhibits sub-gap features originating from multiple Andreev reflections (MARs). MAR features appear at bias voltage values $V_b$ satisfying the condition $V_b=2\Delta/ne$, up to the point when the voltage drop across the junction is $V_b>2\Delta/e$ ($\Delta$ is the superconducting gap and $n$ an integer number). They are a common feature in Josephson junctions with highly transparent interfaces (as reported for example in Ref.~\citenum{Salimian2021}), including Nb-graphene Josephson junctions \cite{BenShalom2016}.

In Fig.~\ref{fig_add_char}a, on the p-type doping side, MARs can be identified as darker stripes corresponding to local maxima in $dV/dI$. They are not observed at fixed $I_{bias}$, because the junction's resistance is not constant but modulated by $V_{BG}$. Selected $dV/dI$ curves as a function of current bias $I_{bias}$ in a $1$~V wide range around the CNP are reported in Fig.~\ref{fig_add_char}b, where the guiding black dashed lines identify the four visible $dV/dI$ peaks due to MARs (n=$1-4$).

A common figure of merit used to assess the quality of a Josephson junction is given by the $I_c\times R_n$ product, which is shown for our device in Fig.~\ref{fig_add_char}c in units of $\Delta/e$. The $I_c\times R_n$ product is almost constant on the n-type doping side, while it is lower on the p-type doping side due to the additional p-n interfaces. The authors of Ref.~\citenum{BenShalom2016} observe that the experimentally obtained value of the $I_c\times R_n$ product is smaller than the theoretical value equal to $\alpha\Delta/e$, where $\Delta$ is the superconducting gap and $\alpha=2.1$.  The $I_c\times R_n$ for our device corresponds to $\alpha\le0.4$, comparable to the value obtained in Ref.~\citenum{BenShalom2016} for a $400$~nm junction with Nb-contacts.

\section{Flux focusing}

\begin{figure}[!htb]
	\centering
	\includegraphics[width=0.6\linewidth]{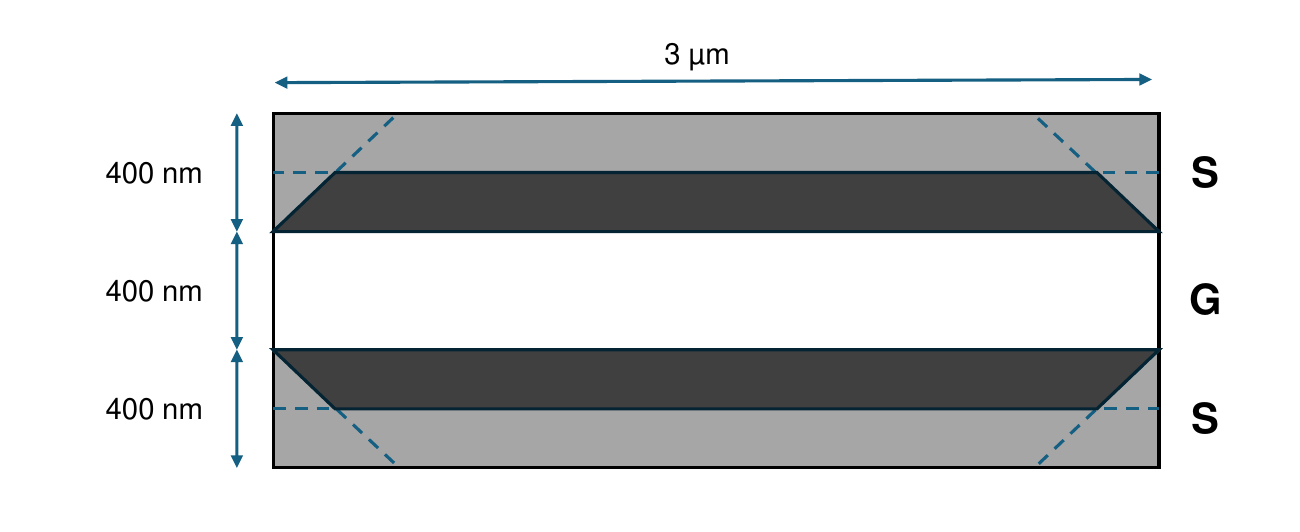}
	\caption{Junction's geometry: the focusing area is calculated as the trapezoidal shape indicated in dark grey in the Nb leads, which corresponds to the Nb lead region which is closer to the Nb-graphene edge than any other edge. S = superconductor. G = graphene. \label{fig_focusing}}
\end{figure}

As reported in the main text, a focusing factor $\simeq 1.7$ is obtained from the position of the first minima  of the Fraunhofer pattern. This is compatible with what we would obtain by estimating the focusing area of the Nb leads (i.e., the area of the leads that expels magnetic field inside the graphene channel). Following the argument reported in Ref.~\citenum{Amet2016}, this can be approximated as the area of the Nb leads that is closer to the graphene channel than every other edge. For our junction's geometry, this corresponds to the trapezoidal shape shown in Fig.~\ref{fig_focusing} in dark grey. Given the Nb leads dimensions ($L=400$~nm, $W=3\;\mu$m), the focusing area for each Nb lead is $A_{foc}=0.55\;\mu$m$^2$, thus the total focusing area $2A_{foc}=1.1\;\mu$m$^2$. This is roughly $90\%$ of the area of the graphene channel, from which we estimate a theoretical focusing factor $\simeq 1.8$, compatible with the value $\simeq 1.7$ extracted from first minima position.

\section{Superconducting pockets in the semiclassical regime}

\begin{figure}[!htb]
	\centering
	\includegraphics[width=\linewidth]{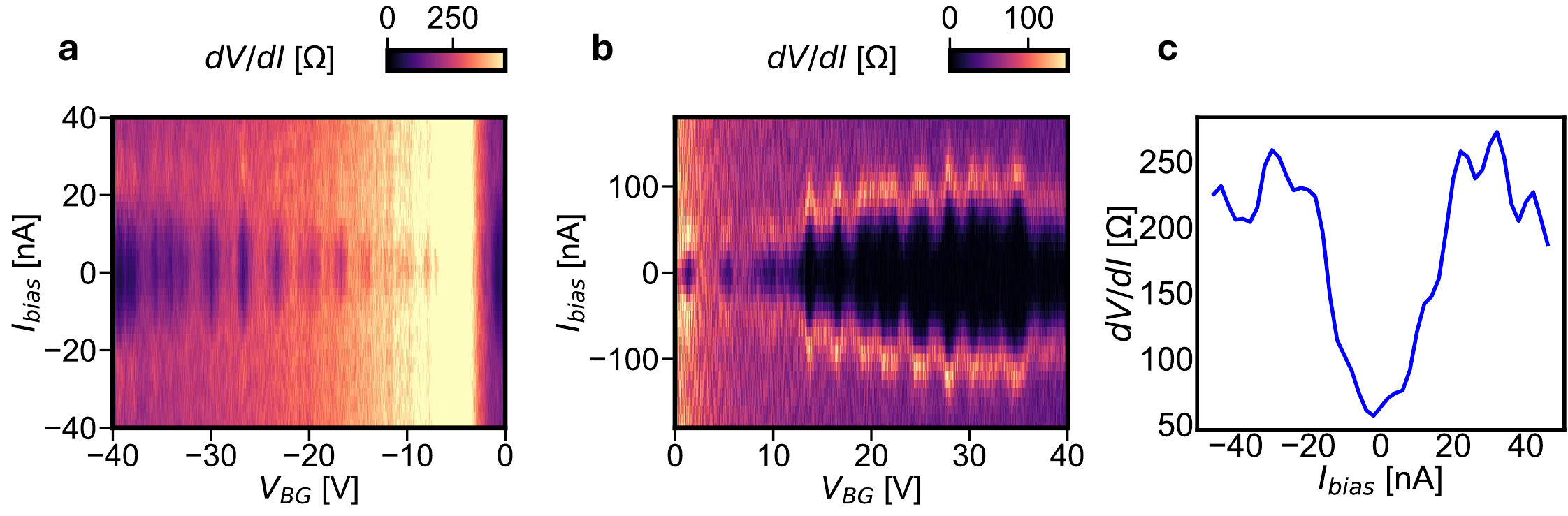}
	\caption{($\mathbf{a}$,$\mathbf{b}$) Differential resistance $dV/dI$ \textit{versus} BG voltage $V_{BG}$ and DC bias current $I_{bias}$. Applied AC bias: $1$~nA. $B=200$~mT, $T=240$~mK. ($\mathbf{c}$) Example  $dV/dI$ linecut ($V_{BG}=-39.3$~V). \label{fig_pockets_semicl}}
\end{figure}

In Fig.~\ref{fig_pockets_semicl} we report the observation of superconducting pockets in the semiclassical regime. Figs. \ref{fig_pockets_semicl}a,b show the differential resistance $dV/dI$ as a function of $V_{BG}$ and $I_{bias}$ at fixed magnetic field, $B=200$~mT. Pockets are observed in both p-type and n-type doping regimes (similar evidence was reported in Ref.~\citenum{Amet2016}, while no traces of supercurrent for p-type doping were reported in Ref.~\citenum{BenShalom2016} for junctions with Nb contacts). A linecut example of a pocket in the p-type doping regime ($V_{BG}=-39.3$~V) is shown in panel \textbf{c}. Compared to the QH regime, pockets with larger $I_c$ are observed (here up to $\sim 100$~nA).

In the semiclassical regime, superconducting states have been attributed to Andreev bound states made of closed trajectories formed by the combination of Andreev reflections at the graphene-superconductor interface and random elastic scattering at the graphene-vacuum interface \cite{BenShalom2016}. This "chaotic ballistic billiard", as defined by the authors, leads to the formation of closed loop trajectories which are capable of transferring Cooper pairs between the superconducting contacts, despite electrons and holes moving along different non-retracing paths.

\section{Temperature dependence of the pocket amplitude in the QH regime}

\begin{figure}[!htb]
	\centering
	\includegraphics[width=0.97\linewidth]{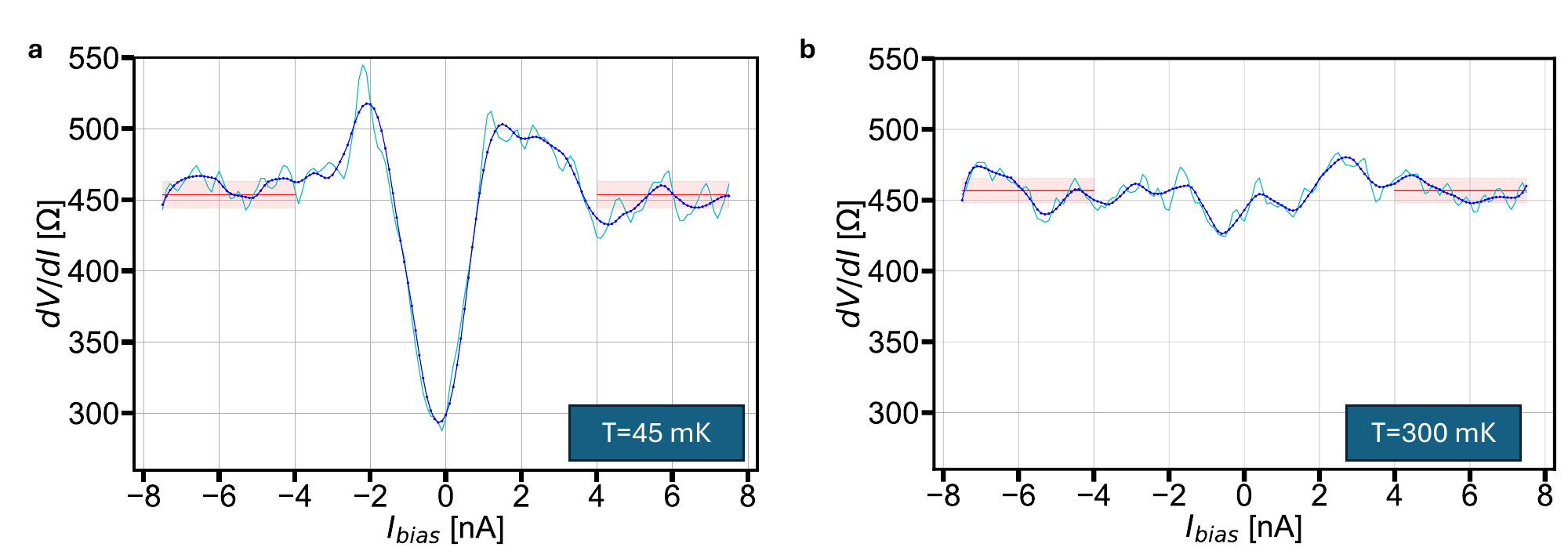}
	\caption{Differential resistance $dV/dI$ \textit{versus} DC bias current $I_{bias}$. Light blue line is the raw data, the blue line is the smoothed data, as explained in the text. Red shaded areas are used to calculate the normal state resistance. ($\mathbf{a}$) Data for $T=45$~mK. ($\mathbf{b}$) Data for $T=300$~mK. \label{fig_pockets_Tdep}}
\end{figure}

\begin{figure}[!htb]
	\centering
	\includegraphics[width=0.97\linewidth]{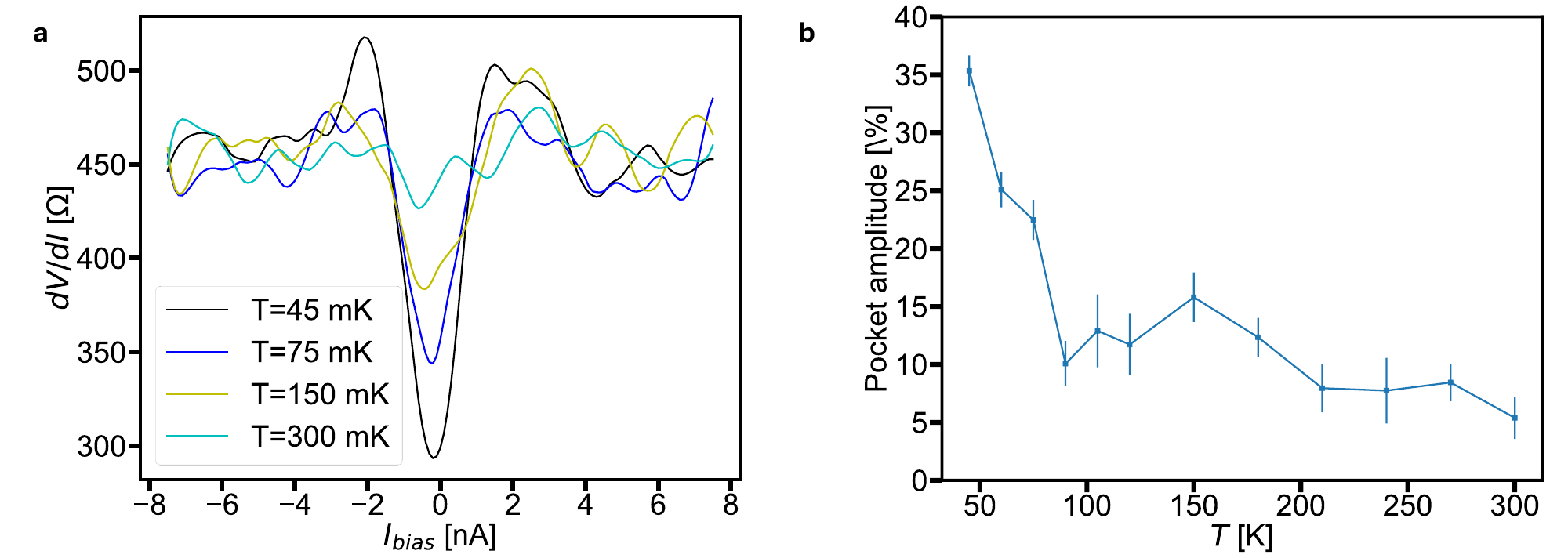}
	\caption{($\mathbf{a}$) $dV/dI$ as a function of the DC bias current $I_{bias}$ for selected temperature values. ($\mathbf{b}$) Pocket amplitude as a function of the temperature, calculated as explained in the text. Errorbars are obtained by taking into account the error on the normal state resistance value (estimated as standard deviation, red shaded regions in Fig.~\ref{fig_pockets_Tdep}). \label{fig_pockets_Tdep_all}}
\end{figure}

Here we discuss the temperature dependence of the amplitude of the superconducting pockets in the quantum Hall regime. The investigated pocket is located at $B=1.6$~T, $V_{BG}=11.6$~V (same starting point for analysis shown in Fig.~4 of main text). We define the amplitude of a pocket as the ratio between the zero-bias $dV/dI$ minimum and the normal state resistance. Based on this definition, a pocket has amplitude $1$ ($100\%$) if $dV/dI=0$ at 0 bias (total suppression of $dV/dI$); on the opposite, in the case of $dV/dI$ at 0 bias equal to the normal state resistance value, no superconducting pocket is present (it has $0$ amplitude). The normal state resistance value is calculated as the mean value of the measured differential resistance at large enough (positive and negative) DC bias ($I_{bias}>4$ nA) to suppress the supercurrent. As an example we report in Fig.~\ref{fig_pockets_Tdep} two acquisitions, at $T=45$~mK (panel \textbf{a}) and $T=300$~mK (panel \textbf{b}). Red lines indicate the calculated mean value of the resistance, while the shaded red areas correspond to the standard deviation of this value.

Prior to calculating the pocket amplitude, a mild smoothing operation was performed (implemented through the \texttt{savgol\_filter} function of \texttt{scipy.signal} Python package \cite{Savitzky1964}, by setting $W=26$ and $n=5$). In Fig.~\ref{fig_pockets_Tdep} both raw (thin light blue continuous line) and smoothed data (blue dot-line) are reported.

The amplitude of the superconducting pocket is rapidly suppressed with increasing temperature, and becomes negligible ($<10\%$) for $T > 200$ mK, as can be seen from Fig.~\ref{fig_pockets_Tdep_all}a. The overall temperature dependence is shown in Fig.~\ref{fig_pockets_Tdep_all}b. As discussed in Ref.~\citenum{Amet2016}, the Josephson energy $E_J=\hbar I_c/2e$, which is of the order of $2\;\mu$eV for supercurrents of $1$ nA, is of the same order of magnitude and very close to the thermal energy scale $k_B T\simeq 2.6\;\mu$eV at $T=30$ mK. By increasing the temperature, the thermal energy rapidly exceeds the Josephson energy, thus destroying the supercurrent in the QH regime.

\section{Coupling of B-modulated supercurrents and sum frequency in the FFT spectrum}

In Fig.~4c of the main text, we show that a fixed-$\nu$ acquisition reveals two main periodicities of the supercurrent pockets, indicating interference in the whole junction area and in the smaller area defined by doping. 
Additionally, a third peak in the FFT spectrum appears at a frequency that quantitatively corresponds to the sum of the two main frequency components. 
The observation of this third peak provides evidence of the coupling of two B-field-modulations of the supercurrent in the two different areas, which we explain in detail in the following.

\begin{figure}[!htb]
	\centering
	\includegraphics[width=0.9\linewidth]{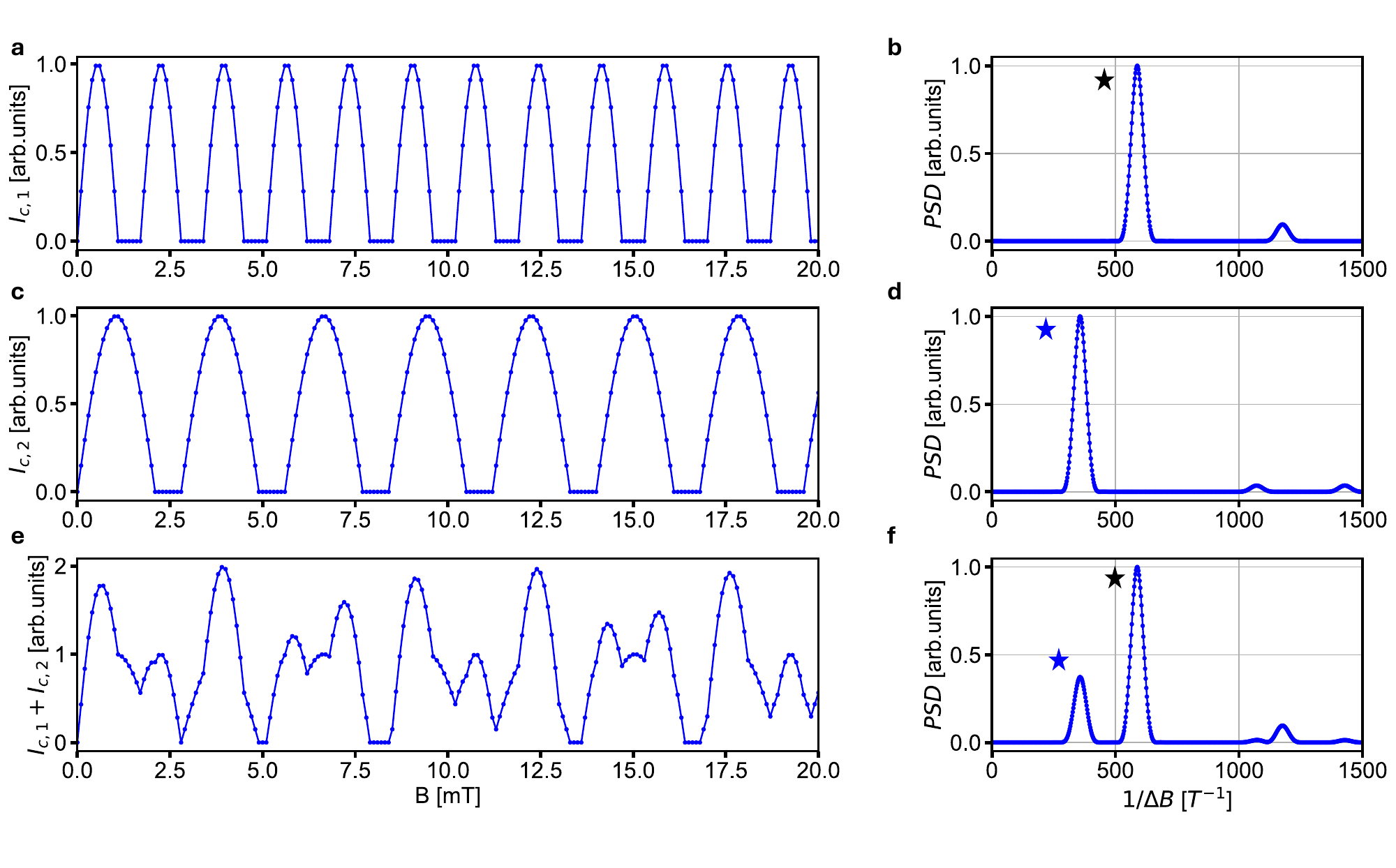}
	\caption{Simulation of independent $B$-modulated supercurrent pockets. ($\mathbf{a}$) $I_{c,1}$ vs.~$B$. ($\mathbf{b}$) Fourier transform (power spectral density) of the data in panel $\mathbf{a}$. The peak indicated with the black star corresponds to the periodicity of the simulated pockets in $\mathbf{a}$. ($\mathbf{c}$) $I_{c,2}$ vs. $B$. ($\mathbf{d}$) Fourier transform (power spectral density) of the data in panel $\mathbf{c}$. The peak indicated with the blue star corresponds to the periodicity of simulated pockets in $\mathbf{c}$. ($\mathbf{e}$) $I_{c,1}+I_{c,2}$ vs. $B$. ($\mathbf{f}$) Fourier transform (power spectral density) of the data in panel $\mathbf{e}$. The peaks indicated with the black and blue stars correspond to the individual periodicities of $I_{c,1}$ and $I_{c,2}$, but the third peak at the sum frequency is absent. \label{fig_Ic_sum}}
\end{figure}

\begin{figure}[!htb]
	\centering
	\includegraphics[width=0.9\linewidth]{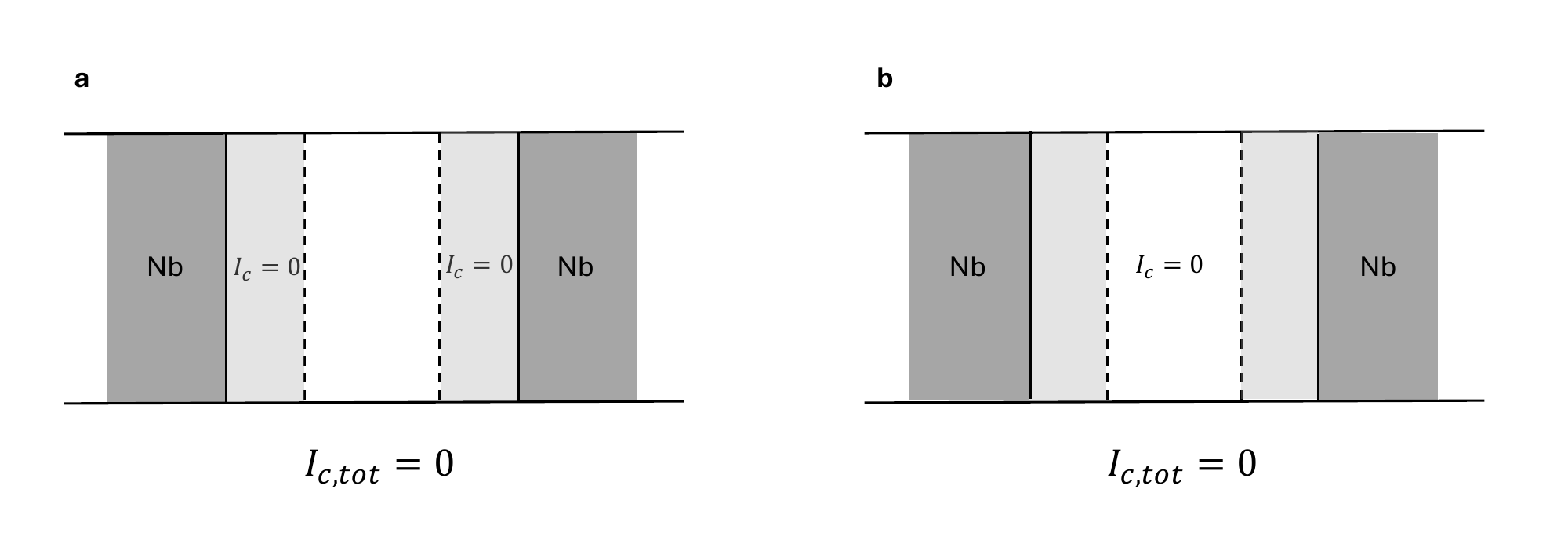}
	\caption{($\mathbf{a}$) Schematics of the junction when destructive interference takes place in the regions close to Nb contacts. ($\mathbf{b}$) Schematics of the junction when destructive interference takes place in the cavity originated by the doping variation. In both cases, the total supercurrent through the junction is equal to 0. \label{fig_schema_correnti}}
\end{figure}

If the two modulations were independent, we would find only two peaks in the FFT spectrum. This case is shown in Fig.~\ref{fig_Ic_sum}, where we simulate two $B$-modulated critical currents ($I_{c,1}$, Fig.~\ref{fig_Ic_sum}a and $I_{c,2}$, Fig.~\ref{fig_Ic_sum}c) and their sum ($I_{c,1}+I_{c,2}$, Fig.~\ref{fig_Ic_sum}e). We consider discrete pockets (i.e., with extended intervals of suppressed supercurrent) as observed experimentally. The non-sinusoidal shape of $I_{c,1}$ and $I_{c,2}$ gives rise to overtones in the FFTs (Figs.~\ref{fig_Ic_sum}b,d), but the third peak at the sum frequency is absent in Fig.~\ref{fig_Ic_sum}f.

In our device, however, the two modulations are not independent. Specifically, if destructive interference takes place in the regions close to the leads, hence the critical current there is 0 (as schematically shown in Fig.~\ref{fig_schema_correnti}a), no supercurrent can flow through the junction. Analogously, destructive interference in the cavity (Fig.~\ref{fig_schema_correnti}b) prevents any supercurrent flow across the device. Indeed, destructive interference over one area necessarily suppresses the supercurrent of the whole junction. In conclusion, when the two B-modulations are not independent, the smallest critical current dominates (which also includes the condition that, if any of the two supercurrents is 0, then the total supercurrent is also 0).

\begin{figure}[!htb]
	\centering
	\includegraphics[width=0.9\linewidth]{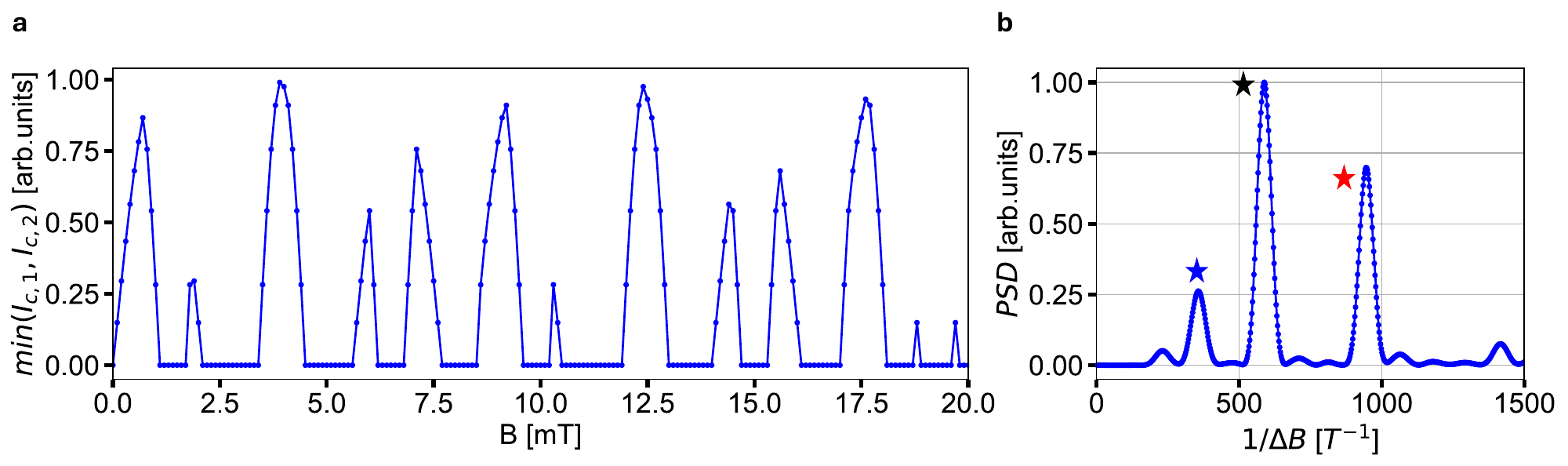}
	\caption{($\mathbf{a}$) Minimum between $I_{c,1}$ and $I_{c,2}$ as a function of $B$. ($\mathbf{b}$) Fourier transform (power spectral density) of data in panel $\mathbf{a}$. Peaks labelled with a black and a blue star correspond respectively to the periodicities of $I_{c,1}$ and $I_{c,2}$, as shown in Fig.~\ref{fig_Ic_sum}. The third peak, labelled with a red star, is located at a frequency corresponding to the sum of the two main oscillatory components. \label{fig_Ic_min}}
\end{figure}

\begin{figure}[!htb]
	\centering
	\includegraphics[width=0.9\linewidth]{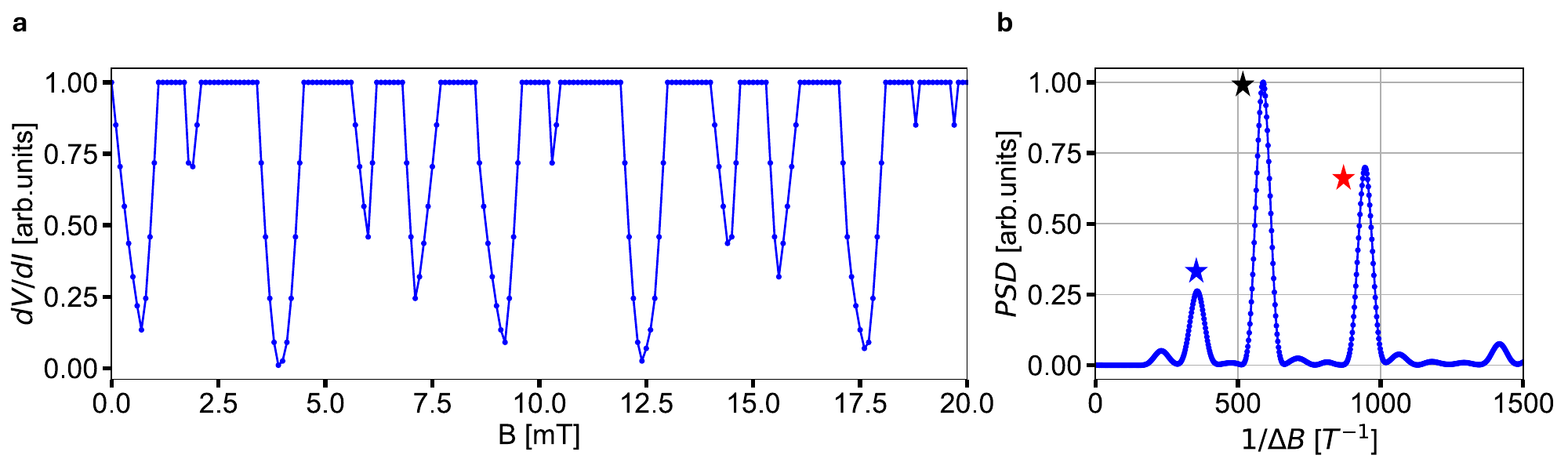}
	\caption{($\mathbf{a}$) Simulation of $dV/dI$ from the critical current shown in Fig.~\ref{fig_Ic_min}a. The differential resistance is suppressed when the critical current is different from zero. ($\mathbf{a}$) Fourier transform (power spectral density) of data in panel $\mathbf{a}$. The FFT spectrum corresponds to the one in Fig.~\ref{fig_Ic_min}b: the third peak is also present. \label{fig_dVdI_PSD}}
\end{figure}

The minimum between $I_{c,1}$ and $I_{c,2}$ and the corresponding FFT spectrum are reported in Fig.~\ref{fig_Ic_min}. The FFT clearly shows a third peak (labelled with a red star), located at the sum frequency of the two fundamental modes.

In the experiment, we rely on the measurement of zero-bias $dV/dI$, which is modulated in the same way as the critical current ($dV/dI$ is suppressed when $I_c\neq 0$). In Fig.~\ref{fig_dVdI_PSD} we show a simulation of $dV/dI$ and the corresponding FFT, again showing the three peaks.

\bibliography{Bibliography_paper_SC-QH_250717_final} 

\end{document}